\documentclass[journal]{IEEEtran} % 2 COL
\usepackage{cite}
\usepackage{latexsym}
\usepackage{verbatim, amsbsy}
\usepackage[dvips]{graphicx}
\usepackage{subfig}
\usepackage{color}
\usepackage{amsmath}                      %!! A lot of other package connected to this must be imported !!
\usepackage{amssymb}
\usepackage{amsthm}
\usepackage{amsfonts}
\usepackage{times}
\usepackage[english]{babel}
\usepackage[dvips]{graphicx}
\usepackage[normalem]{ulem}
\usepackage{bm}
\usepackage{array}
\usepackage{tikz}
\usetikzlibrary{arrows,positioning,shapes.geometric,decorations.pathreplacing,fit}

\usepgflibrary{shapes.geometric}

\newcommand{\ben}{\begin{enumerate}}
\newcommand{\een}{\end{enumerate}}
\newcommand{\be}{\begin{equation}}
\newcommand{\ee}{\end{equation}}
\newcommand{\bea}{\begin{eqnarray}}
\newcommand{\eea}{\end{eqnarray}}
\newcommand{\bc}{\begin{cases}}
\newcommand{\ec}{\end{cases}}
\newcommand{\bi}{\begin{itemize}}
\newcommand{\ei}{\end{itemize}}

\newcommand{\spa}{\,\,\,\!\!}

\newtheorem{teo}{Theorem}[section]
\newtheorem{prop}[teo]{Proposition}

\newtheorem{lem}{Lemma}[section]

 \newtheorem{rem}{Remark}[section]

\def\de{\mathrm{d}}
\def\chitx{\chi_{\rm{tx}}}

\def\chitxT{\chi_{\rm{tx|T}}}
\def\pblo{p_{\rm{blo}}}
\def\ptx{p_{\rm{tx}}}
\def\ptxT{p_{\rm{tx|T}}}
\def\pok{p_{\rm{succ}}}
\def\pdel{p_{\rm{del}}}

\newcommand{\figw}{0.5\columnwidth}
\newcommand{\figL}{0.88\columnwidth}

\begin{document}
\title{Channel, Mode and Power Optimization for non-Orthogonal D2D Communications:\\ a Hybrid Approach}

\author{Federico~Librino,~\IEEEmembership{Member,~IEEE,}
        and~Giorgio~Quer,~\IEEEmembership{Senior Member,~IEEE.}% <-this % stops a space
\thanks{A preliminary and partial version of this paper has been presented at the Information Theory and Applications workshop, San Diego, 2017~\cite{ITA17}.}
\thanks{F. Librino is with the Italian National Research Council, 56124 Pisa, Italy (e-mail: federico.librino@iit.cnr.it).}
\thanks{G. Quer is with the Scripps Research Translational Institute, 3344 North Torrey Pines Court, La Jolla, CA 92037 (e-mail: gquer@scripps.edu).}}

\maketitle

\thispagestyle{empty}

\begin{abstract}
The increasing traffic demand in cellular networks has recently led to the investigation of new strategies to save precious resources like spectrum and energy.
Direct device-to-device (D2D) communication becomes a promising solution if the two terminals are located in close proximity. In this case, the D2D communications should coexist with cellular transmissions, so they must be carefully scheduled in order to avoid harmful interference impacts.
In this paper, we outline a novel framework encompassing channel allocation, mode selection and power control for D2D communications.
Power allocation is done in a distributed and cognitive fashion at the beginning of each time slot, based on local information, while channel/mode selection is performed in a centralized manner only at the beginning of an epoch, a time interval including a series of subsequent time slots.
This hybrid approach guarantees an effective tradeoff between overhead and adaptivity.
We analyze in depth the distributed power allocation mechanism, and we state a theorem which allows to derive the optimal power allocation strategy and to compute the corresponding throughput. Extensive simulations confirm the benefits granted by our approach, when compared with state-of-the-art distributed schemes, in terms of throughput and fairness.
\end{abstract}

\section{Introduction and Related Work}
In the next future, the number of wireless devices is expected to further increase, and unprecedented levels of network density are going to be reached. Furthermore, the amount of traffic exchanged among this huge number of devices is also doomed to raise its growing pace, due to the tremendous demand for high-data rate and low latency wireless services. Since the spectrum resources, although expanded by recent technological advancements, remain limited, the current trend is undoubtedly focused on spectrum reuse. Due to their capability to enhance spatial multiplexing, D2D communications have been envisioned by 3GPP as a promising way to improve network performance~\cite{Rel3GPP,Rel3GPPbis}, and have been deeply investigated in the recent literature~\cite{N1}.

One of the most challenging aspects about D2D communications is the design of efficient and reliable resource allocation schemes, capable of improving the cell capacity without hampering the quality of service (QoS) of authorized users~\cite{N4}.
In current cellular systems, multiple users can perform simultaneous transmissions by exploiting the orthogonal frequency division multiplexing (OFDMA) principle, which prescribes to allocate the available subcarriers to different users. Each subcarrier can be assigned to at most one user, thus avoiding an overwhelming intra-cell interference at the base station (BS).
Conversely, D2D users are distributed in pairs within the cell, and their transmission distance is often much smaller than the cell radius. This potentially allows the reuse of subcarriers among D2D pairs and between D2D users and cellular users, provided that the resulting interference is properly limited.

There are two main approaches to the resource allocation for D2D users, the out-band and the in-band.
According to the \emph{out-band} approach, a dedicated spectrum fraction is reserved to D2D communications only. Authors in~\cite{N2} propose a geographic cell partitioning in order to limit the inter-D2D interference: orthogonal subcarriers are assigned within a subcell, with lower channel state information (CSI) overhead, while power control is designed to maximize the average ergodic capacity.
While ensuring that cellular users do not suffer from additional interference, out-band schemes offer poor spectrum reuse and require the availability of additional subcarriers, which might not be feasible in practical scenarios.
The \emph{in-band} approach instead fully leverages the spectrum sharing principle~\cite{N3} by utilizing the same spectrum for both cellular and D2D communications. Due to its higher potential for intensive spectrum reuse, this approach is currently being deeply investigated. The schemes which follow this line can be further divided into two groups, namely \emph{overlay} and \emph{underlay} schemes.
In overlay schemes~\cite{N7}, D2D communications can take place only on temporarily unoccupied licensed channels, taking advantage of otherwise wasted resources. This ensures that the QoS of cellular users is preserved, and only inter-D2D interference must be managed, without the need for an interference coordination between D2D pairs and cellular users.
In underlay schemes, conversely, D2D devices are allowed to share the spectrum with cellular users, and a licensed channel can be reused among one cellular user and one or more D2D pairs per cell. 
The interference management scheme is of key importance in this case to 
maximize the spectral efficiency while protecting the QoS of the authorized users.
Underlay schemes can attain the highest degree of spectrum sharing, since any channel can be shared by multiple users. In this paper, we follow this approach.
In order to tackle the non trivial problem of interference management, resource allocation and power control play a fundamental role in underlay D2D cellular networks~\cite{N6}. Usually, the condition for a resource sharing is expressed in terms of the resulting signal to interference plus noise ratio (SINR). Unfortunately, given its non convexity and the binary constraints required for channel allocation, the resource allocation problem is often modeled as a non-convex mixed-integer problem, which was proved to be NP-hard~\cite{N12}. Standard Lagrange dual relaxation has been frequently chosen to obtain close-to-optimal solutions~\cite{N20}, but graph theory has recently been leveraged too~\cite{N0,N10,N18}. An interference graph~\cite{N21} is built by assigning to each edge a weight proportional to the interference  across the users represented by the two connected vertices, and allows to turn the optimal allocation problem into a graph coloring problem, which can be solved in polynomial time. Authors in~\cite{N10} extends this concept by creating an hypergraph, which accounts for the fact that interference can come from multiple sources sharing the same channel. A similar approach is described in~\cite{N18}, where the weights of the graph are represented by matrices instead of real numbers, taking into account the different channel responses on different subcarriers. The obtained assignment scheme is shown to approach the optimal solution even for a high number of users.

Besides resource allocation, power control is also pivotal to exploit the spatial dispersion of D2D pairs. Since the D2D link distance is usually short, the required transmission power can be lowered in order to reduce the interference on surrounding ongoing communications. A joint power and resource allocation in multi-tier networks is very challenging~\cite{N5}. Instead, extensive research has been carried out by decoupling the problem into two subproblems or adopting suboptimal heuristics. In~\cite{N8}, authors use graph theory to determine the channel allocation, similar to our approach, but a game theoretic approach is designed to perform power control with incomplete CSI.

Most of the proposed power control and/or resource allocation schemes are obtained as solutions of suitably defined optimization problems. The implementation of such schemes may be centralized or distributed.
Centralized schemes~\cite{resouopt,reallo,N10,N14,N16} aim at collecting all the relevant parameters at a computational entity (usually the BS), including the CSI among all the users, as well as topology and shadowing information. Then a centralized algorithm is run at the BS to determine the optimal solution, which is finally notified to all the users. 
While attaining optimal performance, the major drawback of these schemes lies in the huge level of signaling overhead required for the information gathering. Furthermore, some of the proposed algorithms are not scalable, and their computational burden becomes quickly unsustainable for dense networks, thus making the algorithm hardly implementable in practical systems.

Distributed schemes have also been considered~\cite{N0,N9,N17,modpowcon}. Most of them are based on local information exchange among closely located nodes, which autonomously perform channel selection and/or power adjustment. Some kind of support may be offered by the BS, either feeding back partial CSI or broadcasting updated relevant parameters.

A centralized and a distributed power control scheme in a single cell scenario with multiple D2D pairs are compared in~\cite{N9}. While the centralized algorithm achieves optimality and maximizes the cellular user SINR under constraints for the QoS of D2D communications, it requires global CSI. 
The sub-optimal distributed algorithm with only local CSI signaling is shown to outperform a baseline scheme without D2D communications. 
% \cite{N17} outlines two distributed admission and power control algorithms, aiming at maximizing the capacity of D2D links and the number of D2D pairs when cognitive radio techniques are exploited.
Instead of exchanging several overhead packets to acquire CSI knowledge, authors in~\cite{modpowcon} rely on the topology knowledge for mode selection and power control, thus deriving a lightweight distributed scheme. Despite similar to our approach, however, this scheme lacks the fundamental adaptivity of our framework.

Hybrid schemes have also been investigated. In~\cite{N15}, D2D users determine their transmission power level based on statistical estimates of their channel towards the BS and on SINR degradation margin broadcast by the BS itself. 
\cite{N11} focuses on a single-cell scenario where the mode selection is performed by the BS. D2D users however implement a low-overhead distributed algorithm to determine the access to the resource blocks with varying interference tolerance levels at the BS. A single cell scenario is studied in~\cite{modsel}, too, where a problem-specific Markov Chain is built to maximize the utility of all users, provided that the interference on cellular users remain bounded. The problem of optimal power allocation in underlay/overlay downlink D2D networks is tackled by means of potential games in~\cite{dispow}, which are used to outline two practical iterative algorithms converging to a local sum-rate maximum. As in our paper, some information must be exchanged between terminals or with the BS.
In our previous work~\cite{ourglobe}, we designed a power allocation scheme based only on statistical information. In that work, however, the transmit power level is selected based on the output of Bayesian networks tailored to that scenario, without seeking for an optimal solution.

\subsection{Paper Contributions}
A hybrid approach to the mode, power and channel allocation is pursued in this paper, too. Differently from existing works, where all the information (topology and CSI) is used to determine the optimal or a feasible allocation, we observe that the parameters that influence the signal and interference levels vary at different temporal scales. On one side, topology and shadowing effects are likely to change at a relatively low pace, and might be considered as constant for short periods of time. On the other side, channel conditions, and especially fading coefficients, are likely to vary on a much shorter time scale, and information about them quickly becomes outdated.
Based on this observation, in our work we design a joint centralized mode and channel allocation scheme to tackle the effects of topology and shadowing, while power control is implemented in a distributed way to combat the impairing fading variations.
The operational mode (D2D or cellular mode) and the frequency resources to be assigned at each terminal are derived by the BS on a long time scale, based on the average throughput that can be computed by averaging the fading effects. This allows to strongly reduce the signaling overhead, avoiding the gathering of redundant information across time.
On a much smaller time scale, D2D terminals implement a power control strategy based only on local information and channel statistics, which aims at maximizing a properly defined utility function while limiting the interference on cellular communications. Since the distributed nature of the power control scheme cannot ensure that the interference constraints are constantly matched, a simple reactive protection mechanism is also implemented at the BS to guarantee the QoS of cellular users.

The main contributions of our paper are listed below.

\noindent\textbullet~~ We present a novel cognitive distributed power allocation strategy based only on local CSI for D2D users in an underlay cellular scenario. We state and prove a theorem on the existence and unicity of an optimal strategy that maximizes the utility function defined in terms of average throughput.

\noindent\textbullet~~ We derive the general analytical expressions of the average throughput of a D2D transmitter (DUE) and a cellular user (CUE) sharing the same uplink channel when the derived power control strategy is employed. A closed form expression is obtained for the special case of an on/off power control.

\noindent\textbullet~~ We outline a centralized mode selection and channel allocation scheme, built on top of the abovementioned distributed power control. Channel and mode selection are performed on a much longer time scale, thus strongly reducing the amount of required signaling overhead.
We hence show the benefits, in terms of throughput and fairness, of our hybrid approach with respect to state-of-the-art schemes.

The rest of the paper is organized as follows. In Section~\ref{sec:sysmodel}, we outline our system model, listing the assumptions and the metrics of interest. Section~\ref{sec:opt_mult_act} introduces the utility function, derives the distributed power control strategy maximizing the reward, and proves the unicity of such a strategy.
In Section~\ref{sec:modchasel} we describe the centralized channel and mode selection which leverages the power allocation scheme in order to fully exploit the spectrum resources in the considered cell. We compare the performance of our hybrid approach with that of a distributed strategy in Section~\ref{sec:results}. Section~\ref{sec:conclu} concludes the paper.

\section{System Model}
\label{sec:sysmodel}

\subsection{Primary tier (CUEs)}
We focus on the uplink of a 4G cellular network. We consider a single cell, with $K$ orthogonal uplink channels. We assume full frequency reuse without intra-cell interference, i.e., for each uplink channel $c$, there is exactly one active cellular user (CUE) $U^{(c)}$, transmitting with power $P_{U^{(c)}} = \rho d_{U^{(c)}B}^{\alpha}N_0$. This power is obtained by applying channel inversion, where $N_0$ is the noise power, $B$ is the BS of the considered cell, $d_{XY}$ is the distance between $X$ and $Y$, $\alpha>2$ is the path loss exponent, and $\rho$ is the target SNR of the CUEs.

Time is slotted, and we call an \emph{epoch} a set of $T_e$ subsequent time slots, during which the topology and shadowing effects can be considered as fixed.
All the CUEs are backlogged, i.e., they always have data packets to transmit. All the channels are modeled as Rayleigh channels, meaning that the SINR at $B$ on channel $c$ at time slot $t$ is given by:
\begin{equation}
 \text{SINR}_{B}^{(c)}(t) = \frac{P_{U^{(c)}}d_{U^{(c)}B}^{-\alpha}|h_{U^{(c)}B}^{(c)}(t)|^2}{N_0+I_B^{(c)}(t)} = \frac{\rho|h_{U^{(c)}B}^{(c)}(t)|^2}{1+I_B^{(c)}(t)/N_0} \; ,
\end{equation}
where $h_{X Y}^{(c)}(t)$ is the fading coefficient between $X$ and $Y$ on channel $c$, modeled as a complex Gaussian random variable with zero mean and unitary variance, while $I_B^{(c)}(t)$ is the overall interference perceived at $B$ on channel $c$. In our model, fading coefficients are independent and identically distributed (i.i.d.) across different channels and different time slots.
A data packet sent at time slot $t$ is successfully received at the BS if $\text{SINR}_B^{(c)}(t) > \theta$, where the decoding threshold $\theta$ is a system parameter.

\subsection{Secondary tier (DUEs)}
We consider now another set of backlogged\footnote{The scenario with backlogged sources is the one with the highest amount of traffic. Other traffic models could be incorporated by properly combining the packet generation probability with the channel statistics.} terminals, $\mathcal{S}$, randomly deployed within the cell, which we call DUEs. Each DUE $S \in \mathcal{S}$, has data to transmit to another terminal $D$, located within a distance $d_L$, thus creating a DUE pair.
The DUEs also transmit on the uplink channels, thus sharing them with the CUEs.
Each uplink channel can be shared by at most 1 CUE and 1 DUE, and can be shared either orthogonally or non orthogonally.
In the former case (D2B mode), the DUE $S$ and the CUE $U$ alternatively transmit to the BS on the same uplink channel by adopting a Time Division Multiple Access (TDMA) strategy. Since $S$ now behaves as a normal CUE, it also sets its target SNR to $\rho$. No cross-tier interference is experienced in this case. 
In the latter case (D2D mode), $S$ transmits directly to $D$, and can choose its transmit power among a set $\mathcal{P}$ of available power levels. Here, the cross-tier interference between the communication from $S$ and the communication from $U$ must be limited by a proper power allocation strategy at $S$.

\begin{figure}
 \begin{center}    
 \includegraphics[width=\figL,natwidth=1011,natheight=658]{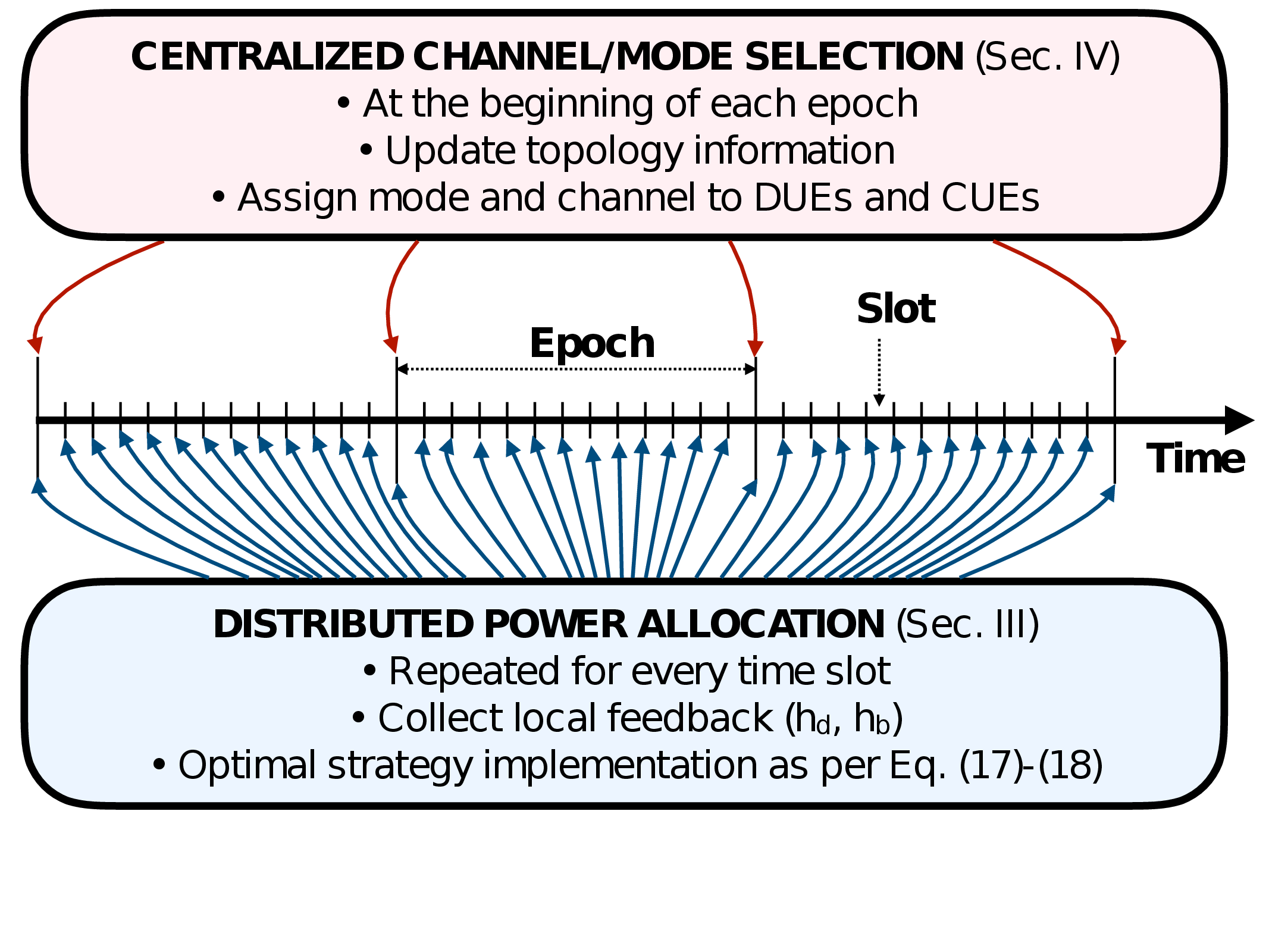}
 \caption{\small Conceptual scheme of our approach. Mode selection is performed only at the beginning of each epoch; power allocation (for D2D mode only) is performed distributedly at the beginning of each time slot.}
 \label{diag:concept}
 \vspace{-1cm}
 \end{center}
\end{figure}

In this paper, we devise a novel channel/mode/power selection approach capable to attain a high aggregated throughput with limited channel information. Differently from fully centralized system, where the entire resource allocation is computed at the BS, based on updated information, we split it as follows.
\begin{itemize}
 \item Channel and mode selection are performed by the BS, relying only on quasi static or stochastic information, at the beginning of each epoch.
 \item Power allocation is instead performed distributedly by each DUE in D2D mode at the beginning of each time slot, based on local instantaneous information.
\end{itemize}
Our hybrid centralized/distributed approach aims at avoiding excessive overhead while keeping the advantages of adaptivity. Fast changing information, like the instantaneous interference level or the fading coefficients, does not need to be gathered at the BS, but is exploited locally for power allocation. Conversely, static or stochastic information, which must be updated less frequently, is leveraged for channel and mode selection in a centralized manner to take advantage of the network infrastructure. A conceptual representation of our scheme is illustrated in Figure~\ref{diag:concept}.

For channel and mode selection, we assume that the cell topology is known at the BS at the beginning of each epoch.
For the power allocation of a DUE $S$ in D2D mode on uplink channel $c$, we assume that the following information is available at $S$ at the beginning of each time slot:
\begin{itemize}
 \item its own location, as well as the one of its intended receiver $D$, of the BS $B$ and of the CUE $U^{(c)}$ transmitting on the same channel;
 \item the perceived interference $I_D(t)$ at $D$, which is fed back on a common out-of-band control channel $c_0$. Since the uplink channels are orthogonal, this is equivalent to know the fading coefficient $h_{U^{(c)}D}(t)$;
 \item the current fading coefficient $h_{U^{(c)}B}(t)$, overheard on the downlink control channel from $B$ to $U^{(c)}$. 
\end{itemize}

Since power allocation is performed distributedly by each DUE in D2D mode, we add a \emph{protection} mechanism to ensure that the CUE transmissions are not impaired by excessive interference. 
On a channel $c$ where there is an ongoing D2D transmission, if $\text{SINR}_{B}^{(c)}(t)$ falls below a warning threshold $\vartheta$, the BS broadcasts a warning message, and the DUE transmitting on channel $c$ is forced to remain silent for the subsequent $W$ time slots.
The warning threshold $\vartheta$ and the blockage duration $W$ are fixed system parameters, which are tuned to balance the tradeoff between the advantage of offloading data transmissions and the disadvantage of an increased interference level. Without loss of generality, we set $\vartheta=\theta$ (the decoding threshold) in the rest of the paper.

\subsection{Problem Formulation}
The usage of the blockage period allows the BS to intervene only when necessary, instead of proactively scheduling all the users at every time slot. In this sense, our approach can be considered a \emph{reactive} scheme. Under these conditions, it is up to the DUE to properly choose its power allocation strategy. The identification of the optimal power allocation strategy in this scenario is the main contribution of the paper, as described in Section~\ref{sec:opt_mult_act}. The centralized channel and mode selection are built on top of this result, and will be outlined in Section~\ref{sec:modchasel}.

The two steps have different purposes. The optimal power allocation strategy $\mu^*_{S,U}$, distributedly employed by a DUE $S$ sharing the channel with a CUE $U$, is the one that maximizes the average DUE throughput $\tau$, namely:
\begin{equation}
 \mu^*_{S,U} = \arg\max_{\mu}\tau\left(\mu_{S,U}\right).
 \label{probform}
\end{equation}
This maximization is not straightforward. We tackle the problem by finding the strategy which maximizes a properly defined reward function $C_{\lambda}(\mu_{S,U})$, parameterized by a real positive number $\lambda$. We then derive the DUE throughput obtained with this strategy as a function of the same parameter $\lambda$, and we finally identify the value $\lambda^*$ which maximizes it.

The centralized mode and channel selection instead aims at pairing the DUEs with the available channels, and hence with the CUEs, with the aim of maximizing the total cell throughput, given by the sum of the throughput values of all the UEs (both CUEs and DUEs):
\begin{eqnarray}
 \mathbf{\Lambda}^* \!\!\!\!& = \!\!\!\!& \arg\max_{\mathbf{\Lambda},\mathbf{x}}\sum_{i=1}^N\left(\mathbf{x}(i)\left(\tau(\mu^*_{\mathbf{\Lambda}(i),U_i})+ \sigma(\mu^*_{\mathbf{\Lambda}(i),U_i})\right)+\right. \nonumber \\
 \!\!\!\!& \!\!\!\!&  + \left(1-\mathbf{x}(i)\right)\nu(\mathbf{\Lambda}(i),U_i)\Big)\!,
 \label{probform2}
\end{eqnarray}
where $N$ is the number of CUEs (and channels); $\mathbf{\Lambda}$ is the $N\times1$ pairing vector, such that $\mathbf{\Lambda}(i)$ is the DUE assigned to CUE $U_i$; $\mathbf{x}$ is the $N\times1$ binary mode vector, such that $\mathbf{x}(i)=1$ if D2D mode is employed on the channel of CUE $U_i$ and 0 otherwise. Moreover, $\sigma(\mu_{S,U})$ is the average throughput of CUE $U$ that shares a channel with DUE $S$ using power allocation strategy $\mu_{S,U}$; and $\nu(S,U)$ is the overall throughput that can be attained by $S$ and $U$ sharing the same channel in an orthogonal way (D2B mode).

\section{Distributed Power Allocation}
\label{sec:opt_mult_act}
A DUE $S$ in D2D mode, which is allocated on channel $c$ shared with CUE $U$, autonomously performs power control at each time slot. We assume that a set $\mathcal{P}$ of $N$ logarithmically spaced power levels is available: $\mathcal{P} = \{P_i = 2^{i-1}P_S\}$, for $i=1,2,\ldots N$, where $P_S$ is a fixed minimum trasmit power level.

At the beginning of each time slot $t$, $S$ obtains information about the current fading coefficients $h_d = |h_{UD}(t)|^2$ and $h_b = |h_{UB}(t)|^2$, which are considered to be fixed for the entire time slot: the couple $(h_d, h_b)$ defines the current \emph{system state}. $S$ selects the power level for slot $t$ in $\mathcal{P}$ based on the system state and on the available topology information. We also consider admissible the selection of a zero power level, meaning that no transmission at all is performed in the current time slot.
Note that the distributed implementation of this closed loop power control requires the exchange of only two feedback information packets per slot. Computing the power control at the BS would instead require at least three transmissions, in order to measure the CSI of the channel between $S$ and $D$, report its value to $B$ and finally notify $S$ with the selected transmit power level.
The overhead reduction would be even higher in a more general scenario, where multiple DUEs are allowed on the same channel, since the information about the interference level at the BS could be broadcast to all the D2D transmitters with one transmission.

\subsection{Power allocation strategies}
A power allocation strategy $\mu:\mathbb{R}^2\rightarrow\{0,1,2,\ldots,N\}$ is a function which maps any possible state $(h_d, h_b)$ into an integer value $i$, where $i=0$ corresponds to no transmission and $1\leq i\leq N$ means that power level $P_i$ is chosen\footnote{With respect to Eq. (\ref{probform}), in this section we drop the subscripts $S$ and $U$ for the sake of clarity, since we focus on a generic CUE-DUE pair.}.
The instantaneous throughput attained by $\mu$ at state $(h_d, h_b)$ is given by the packet decoding probability at $D$, $p_i(h_d) = \mathbb{P}[\textrm{SINR}_D\geq\theta|h_d,i]$. The probability of triggering a blockage is given by $q_i(h_b) = \mathbb{P}[\textrm{SINR}_B<\theta|h_b,i]$. For $i=0$, they are both set to 0. If instead $i>0$, they are expressed as
\begin{eqnarray}
 p_i(h_d) & = & \mathbb{P}\left[\frac{2^{i-1}P_Sd_{SD}^{-\alpha}|h_{SD}(t)|^2}{N_0+P_Ud_{UD}^{-\alpha}h_d}\geq\theta\right] \nonumber \\
 & = & \exp\left(-\frac{\theta(\gamma_{UD}h_d+1)}{2^{i-1}\gamma_{SD}}\right)\;,
 \label{mulpi}\\
 q_i(h_b) & = &  \mathbb{P}\left[\frac{P_Ud_{UB}^{-\alpha}h_b}{N_0 + 2^{i-1}P_Sd_{SB}^{-\alpha}|h_{SB}(t)|^2} < \theta
\right] \nonumber \\
 & = & \min\left[\exp\left(-\frac{\gamma_{UB}h_b-\theta}{\theta2^{i-1}\gamma_{SB}}\right),1\right]\;,
\label{mulqi}
\end{eqnarray}
where $\gamma_{XY} = P_Xd_{XY}^{-\alpha}/N_0$. The expressions in (\ref{mulpi}) and (\ref{mulqi}) are derived considering that the unknown fading coefficients $|h_{SD}(t)|^2$ and $|h_{SB}(t)|^2$ are exponentially distributed with unitary mean, since $h_{SD}(t)$ and $h_{SB}(t)$ are complex Gaussian random variables with zero mean and unitary variance.

A tradeoff occurs in defining a power allocation strategy: selecting low transmit power values reduces the decoding probability, and hence the throughput, while high values can trigger the blockage mechanism, thus forcing $S$ to defer further transmissions by $W$ time slots.
We tackle the problem of finding the optimal strategy, which maximizes the expected throughput, by splitting it into two subproblems. \textbf{(S1)} We first define a utility function that penalizes the occurrence of a blockage by a fixed weight $\lambda$, and derive the strategy which maximizes the expected value of this utility function for a given value of $\lambda$.
\textbf{(S2)} Subsequently, we determine the expression of the throughput achieved by such a strategy as a function of $\lambda$, and we find the value $\lambda^*$ which maximizes the throughput.

\subsection{Utility Function}
To evaluate the performance of a power allocation strategy, we associate to each strategy $\mu$ the utility function $r_{\mu}:\mathbb{R}^+\times\mathbb{R}^+\rightarrow\mathbb{R}$, defined as
\begin{equation}
 r_{\mu}(h_d, h_b) = p_j(h_d) - \lambda q_j(h_b)\;,
\label{defuti}
\end{equation}
with $j={\mu(h_d,h_b)}$ and $\lambda\in\mathbb{R}^+$. At each state $(h_d, h_b)$ of the system, $r_{\mu}(h_d, h_b)$ is equal to the instantaneous throughput obtained with power level $j$, minus a penalty factor proportional to the blockage probability obtained with the same power level. The blockage weight $\lambda$ is a tunable parameter which quantifies the impairment due to a blockage period.

The overall reward $C_{\lambda}(\mu)$ of the strategy is obtained by averaging $r_{\mu}(h_d,h_b)$ over the state probability distribution. Since $h_b$ and $h_d$ are i.i.d. exponential random variables with unitary mean, we have
\begin{equation}
 C_{\lambda}(\mu) = \int_0^{+\infty}\int_0^{+\infty}r_{\mu}(x,y)e^{-x}e^{-y}\de x\de y\;.
 \label{defrew}
\end{equation}
Notice that this value holds for any time slot, due to the time-invariance of the fading coefficients.

\subsection{Maximum Reward power allocation strategy}
For a given value of $\lambda$, the Maximum Reward (MR) power allocation strategy is $\mu_{\lambda}^* = \arg\max_{\mu}C_{\lambda}(\mu)$, i.e., the one that maximizes the reward (\ref{defrew}). We can find it by maximizing the utility function for any value of $(h_d, h_b)$. From (\ref{defuti}) this means that, for each state, we have
\begin{eqnarray}
\mu_{\lambda}^*(h_d, h_b) & = & \arg\max_{\mu}r_{\mu}(h_d,h_b)\nonumber\\
 & = & \arg\max_{i\in\{0,1,\ldots, N\}}\left(p_i(h_d)-\lambda q_i(h_b)\right)\;,
 \label{mulopt}
\end{eqnarray}
where $p_i(h_d)$ and $q_i(h_b)$ are derived in (\ref{mulpi}) and (\ref{mulqi}).
We can distinguish two cases.

\emph{Case 1}, $h_b<\theta/\gamma_{UB}$. In this case, the SNR of the signal from $U$ at the BS is already below the threshold $\theta$, and any transmission from $S$ triggers the blockage. In fact, from (\ref{mulqi}) we get that $q_i(h_b)=1$, $\forall i>0$, and it follows from (\ref{defuti}) that in this case the expected reward using power level $i$ is $p_i(h_d)-\lambda$, if $i>0$, and 0 otherwise.
Since $p_j(x)>p_i(x)$ for $j>i$, $\forall x$, the expected reward is maximized by putting $i = N$, if $p_N(h_d) > \lambda$, and $i = 0$ otherwise. Therefore, using the expression of $p_N(\cdot)$ in (\ref{mulpi}), when $h_b<\theta/\gamma_{UB}$ the MR strategy is defined as
\begin{equation}
 \mu_{\lambda}^*(h_d, h_b) = \begin{cases}
                N & \text{for } h_d < h^* \\
                0 & \text{otherwise}\;,
               \end{cases}
\label{optlowhb}
\end{equation}
where
\begin{equation}
 h^* = -\frac{2^{N-1}\gamma_{SD}}{\theta\gamma_{UD}}\ln(\lambda) -\frac{1}{\gamma_{UD}}\;.
 \label{defhstar}
\end{equation}
Notice that if $\lambda>e^{-\theta/(2^{N-1}\gamma_{SD})}$, then $h^*<0$, and $\mu_{\lambda}^*(h_d,h_b)=0$, $\forall (h_d,h_b):h_b<\theta/\gamma_{UB}$.

\emph{Case 2}, $h_b\geq\theta/\gamma_{UB}$. In this case, the expression of $q_i(h_b)$ is an exponential, as per (\ref{mulqi}). Under this condition, the following lemma holds.
\begin{lem}
There exists a linear function $g_0(h_d) = M h_d + Q$ such that the MR power allocation strategy is $\mu_{\lambda}^*(h_d,h_b)=0$ if $h_b<g_0(h_d)$, and $\mu_{\lambda}^*(h_d,h_b)>0$ otherwise. 
\label{lem:binary}
\end{lem}
\begin{proof}
From (\ref{mulopt}), we have that $\mu_{\lambda}^*(h_d, h_b) = 0$ if $p_i(h_d)-\lambda q_i(h_b)<0$, $\forall i>0$. By replacing the expressions of $p_i(h_d)$ and $q_i(h_b)$ in (\ref{mulpi}) and (\ref{mulqi}), this inequality can be rewritten as
\begin{eqnarray}
  h_b &\!\!\!\! < &\!\! \theta^2\frac{\gamma_{SB}\gamma_{UD}}{\gamma_{SD}\gamma_{UB}}h_d + \theta\frac{\gamma_{SB}}{\gamma_{UB}}\left(\frac{1}{\gamma_{SB}}+\frac{\theta}{\gamma_{SD}} + 2^{i-1}\ln(\lambda)\right) \nonumber \\
  &\!\!\!\! < &\!\! Mh_d + \nu_i,\quad \forall i\in(1,2,\ldots,N)\;.
\end{eqnarray}
Since the above inequality must hold $\forall i>0$, we can state that $\mu_{\lambda}^*(h_d, h_b) = 0$ when
\begin{equation}
 h_b < Mh_d + \min_i(\nu_i)\;.
\end{equation}
This proves the lemma, with $M = \theta^2\gamma_{SB}\gamma_{UD}/(\gamma_{SD}\gamma_{UB})$, while $Q = \nu_1$ if $\lambda\geq1$, and $Q = \nu_N$ otherwise.
\end{proof}

The obtained results for Case 1 and Case 2 can be shown to be consistent. When $\lambda>1$, it follows from (\ref{optlowhb}) that $\mu_{\lambda}^*(h_d, h_b) = 0$, $\forall h_b<\theta/\gamma_{UB}$: coherently, we also have $g_0(h_d)>\theta/\gamma_{UB}$, $\forall h_d\in\mathbb{R}^+$. Conversely, when $\lambda<1$, we find that
$g_0(h_d)$ intersects the line $h_b=\theta/\gamma_{UB}$ exactly for $h_d = h^*$, with $h^*$ defined in (\ref{defhstar}).
This means that, by combining the result from Lemma~\ref{lem:binary} with that in (\ref{optlowhb}), we can state that $\mu_{\lambda}^*(h_d, h_b) = 0$ when

\begin{equation}
 h_b < \tilde{g}_0(h_d) = g_0(h_d)\mathbb{S}\left(h_d-h^* \right)\;,
 \label{defA0}
\end{equation}
being $\mathbb{S}(\cdot)$ the Heaviside step function.

A graphical representation of the MR power allocation strategy can be drawn on the cartesian plane $h_d-h_b$. We call $\mathcal{Z}$ the set of all points with non negative coordinates in this plane, that is, the set of all the system states. $\mathcal{Z}$ can be partitioned into $\mathcal{Z}_0$, $\mathcal{Z}_1$, $\ldots$, $\mathcal{Z}_N$, where $\mathcal{Z}_i$ is the subset of points $(h_d,h_b)$ for which $\mu_{\lambda}^*(h_d,h_b)=i$.
As shown in Lemma \ref{lem:binary}, curve $\tilde{g}_0(h_d)$ represents the upper border of region $\mathcal{Z}_0$.
For the remaning regions, we can state the following lemma:
\begin{lem}
For every $1\leq i<N$, the region $\mathcal{Z}_i$ can be adjacent only to the regions $\mathcal{Z}_{i-1}$ and $\mathcal{Z}_{i+1}$. 
\label{lem:many_regions}
\end{lem}

\begin{proof}
See Appendix~\ref{app:prooflem}.
\end{proof}
$\mathcal{Z}_0$ is not included in the Lemma~\ref{lem:many_regions} since, as seen in the proof of Lemma~\ref{lem:binary}, $\mathcal{Z}_0$ is adjacent only to region $\mathcal{Z}_1$ if $\lambda\geq1$, and only to $\mathcal{Z}_N$ otherwise.

The boundary between two adjacent regions $\mathcal{Z}_i$ and $\mathcal{Z}_{i+1}$, for $i\in\{1,2,\ldots,N-1\}$ can be calculated by solving the equation $p_i(h_d)-\lambda q_i(h_b) = p_{i+1}(h_d)-\lambda q_{i+1}(h_b)$. If $\lambda\geq1$, the equation is solved by the points belonging to two functions
\begin{eqnarray}
 g_i^{\pm}(h_d) &\!\! = &\!\!\!\! \frac{\theta}{\gamma_{UB}}-\theta\frac{\gamma_{SB}}{\gamma_{UB}}2^i\times\nonumber\\
 &\hspace{-3cm} & \hspace{-1.53cm} \times \ln\!\left(\!\frac{1}{2}\mp\frac{1}{2}\sqrt{1-\frac{4}{\lambda}p_{i+1}(h_d)\left(1-p_{i+1}(h_d)\right)}\!\right).
\end{eqnarray}
However, it can be shown that $g_i^-(h_d)<\tilde{g}_0(h_d)$, $\forall h_d\in\mathbb{R}^+$, thus entirely inside $\mathcal{Z}_0$. The only boundary is hence given by $g_i^+(h_d)$.
Similarly, when $\lambda<1$, we derive the two curves
\begin{eqnarray}
 f_i^{\pm}(h_b) &\!\! = &\!\!\!\! -\frac{1}{\gamma_{UD}} - \frac{\gamma_{SD}}{\theta\gamma_{UD}}2^i\times\nonumber \\
 &\hspace{-3cm} & \hspace{-1,5cm}\times \ln\!\left(\!\frac{1}{2}\pm\frac{1}{2}\sqrt{1-4\lambda q_{i+1}(h_b)\left(1-q_{i+1}(h_b)\right)}\right).
\end{eqnarray}
Here, we find that $f_i^-(h_b)>\tilde{f}_0(h_b)$, $\forall h_b\in\mathbb{R}^+$, where $\tilde{f}_0(h_b) = \max((h_b-\nu_0)/m, h^*)$ is the inverse function of $\tilde{g}_0(h_d)$ defined in (\ref{defA0}), with $h^*$ given by (\ref{defhstar}). Therefore, $f_i^-(h_b)$ lies within region $\mathcal{Z}_0$, and the only valid boundary is $f_i^+(h_b)$.

We can state the following remarks about the boundary functions $g_i^+(h_d)$ and $f_i^+(h_b)$:
\begin{rem}
 $g_i^+(h_d)\cap g_j^+(h_d)=\emptyset$, and $f_i^+(h_b)\cap f_j^+(h_b)=\emptyset$, $\forall i\neq j$.
 \label{rem:noint}
\end{rem}
The intersection between $g_i^+(h_d)$ and $g_j^+(h_d)$ must be empty. We can see it by observing that otherwise there would be a point $(h_d,h_b)$ where the maximum reward can be reached using more than two power levels, which is not possible, due to the characteristics of the reward functions $v_{h_d,h_b}(x)$ detailed in the proof of Lemma \ref{lem:many_regions}.

\begin{rem}
 When $\lambda\geq1$, $g_i^+(h_d)\cap\mathcal{Z}\neq\emptyset$, $\forall i\in\{1,2,\ldots,N\}$.
 \label{rem:exig}
\end{rem}
This follows from the fact that $g_i^+(h_d)>g_0(h_d)$, $\forall h_d\in\mathbb{R}^+$, which can be proven through calculations, since $g_0(h_d)$ is an increasing linear function. Note that this implies the existence within $\mathcal{Z}$ of all the regions $\mathcal{Z}_i$, when $\lambda\geq1$.
The same does not hold when $\lambda<1$.

\begin{rem}
 For $\lambda\geq1$, $g_i^+(h_d) > g_j^+(h_d) > g_0(h_d)$, $\forall h_d\in\mathbb{R}^+$, $\forall i>j$. Similarly, for $\lambda<1$, $f_0(h_b) > f_i^+(h_b) > f_j^+(h_b)$, $\forall h_b\in\mathbb{R}^+$, $\forall i>j$.
 \label{rem:order}
\end{rem}
It is not immediate to demonstrate these inequalities via algebraic derivation. We observe that each function $g_i^+(h_d)$, with $i\geq1$, for $h_d\rightarrow\infty$ approaches asymptotically the linear function
\begin{equation}
 \bar{g}_i(h_d) = \theta^2\frac{\gamma_{SB}\gamma_{UD}}{\gamma_{SD}\gamma_{UB}}h_d + \theta\frac{\gamma_{SB}}{\gamma_{UB}}\left(\frac{1}{\gamma_{SB}}+\frac{\theta}{\gamma_{SD}}\! +\! 2^i\ln(\lambda)\right) .
\end{equation}
When $\lambda\geq1$, from the fact that $\bar{g}_i(h_d) > \bar{g}_j(h_d)$, $\forall h_d\in\mathbb{R}^+$ for any $i>j$, it follows that $\exists H\in\mathbb{R}^+:\forall h_d>H\,, \, g_i^+(h_d)>g_j^+(h_d)$. Since $g_i^+(h_d)$ and $g_j^+(h_d)$ never intersect, we obtain the statement in Remark \ref{rem:order}.
Proving the same about the functions $f_i^+(h_b)$ is more involved, but can be done by computing the intersections between each $h_i^+(h_b)$ and a properly chosen linear function $h_d = H$, and verifying how these points are sorted.

Using the previous Lemmas and Remarks, we can state the following Theorem.
\begin{teo}
Given a set of $N$ positive and logarithmically spaced power levels such that $\mathcal{P} = \{0\}\cup\{P_i=2^{i-1}P_S\}$, with $i\in\{1,2,\ldots,N\}$, it is always possible to divide the space $(h_d,h_b)$ into at least 2 and at most $N+1$ continuous regions, such that the MR policy $\mu_{\lambda}^*(h_d,h_b)$ is always unambiguously defined, with the exception of the boundaries between the regions, which have measure zero.
 \label{teo:optsol_multiple}
\end{teo}

\begin{proof}
See Appendix~\ref{app:proofteo}.
\end{proof}

\begin{figure}
    \centering
     \subfloat[$\lambda = 0.8212<1$] %[$K = 1.2177$]
     {\includegraphics[width=\figw]{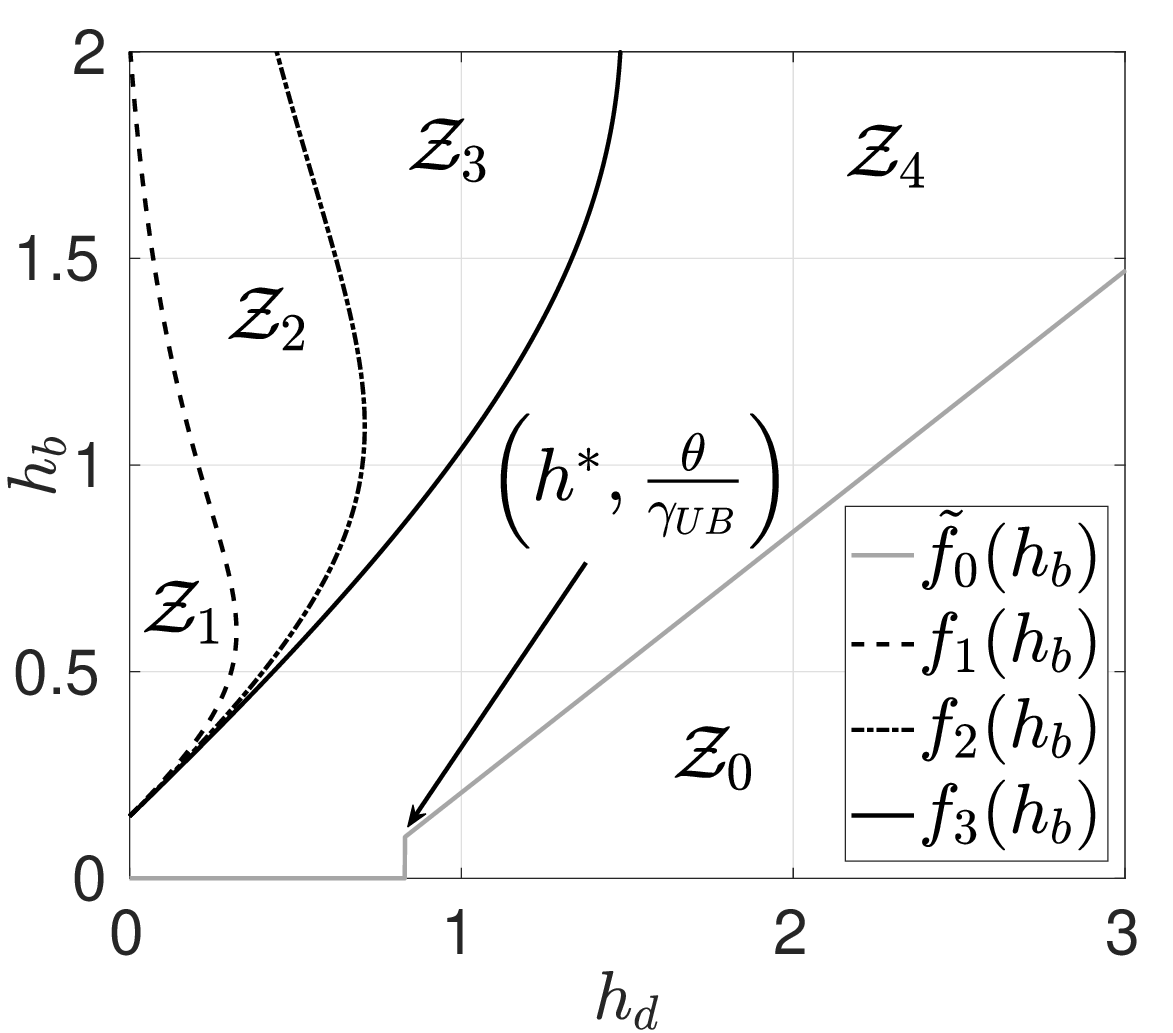}
     \label{fig:opt_multi_khigh}
     }
    \subfloat[$\lambda = 1.1918>1$] % [$K = 0.8391$]
    {\includegraphics[width=\figw]{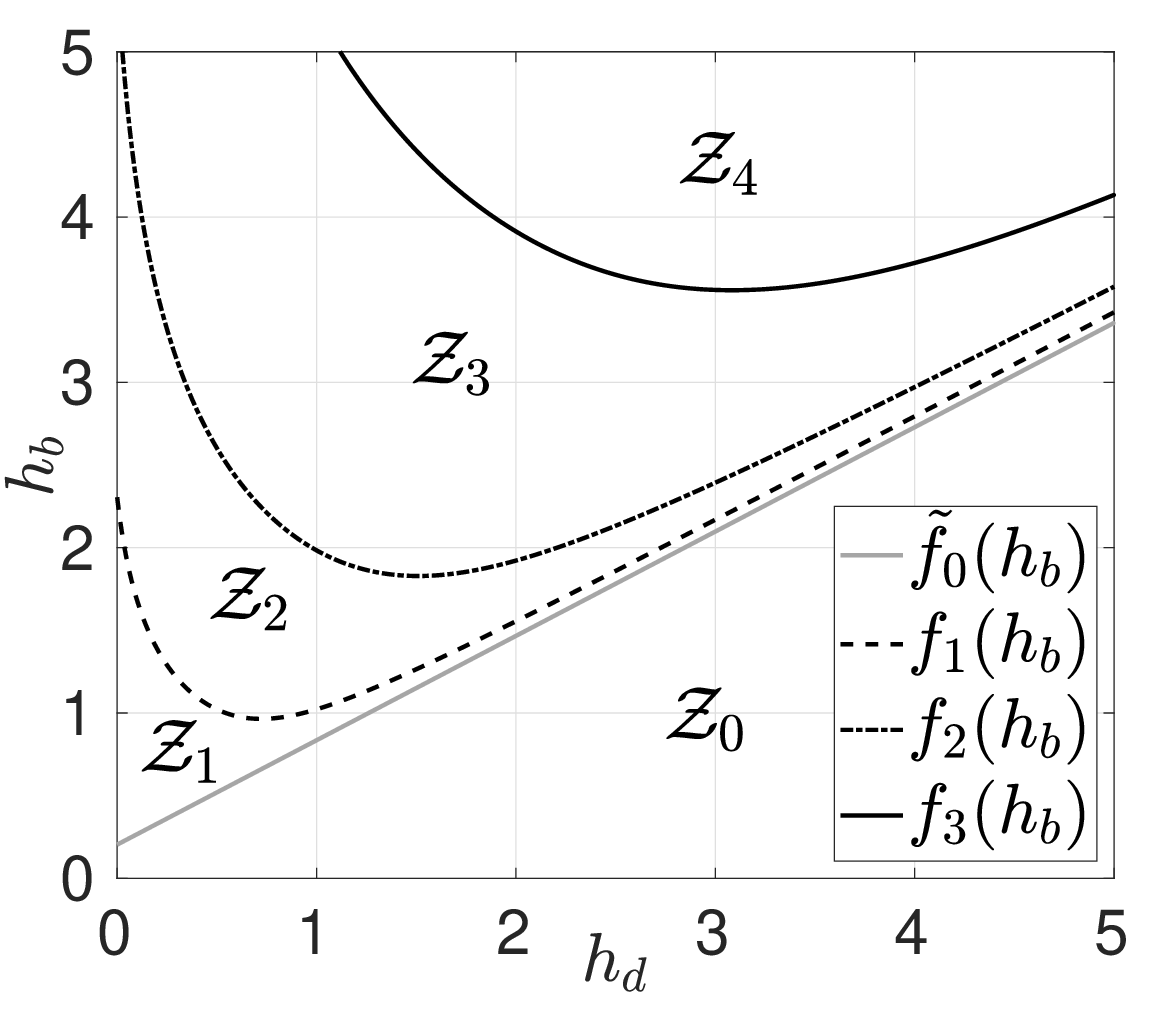}
     \label{fig:opt_multi_klow}
     }     
     \caption{\small The MR policy for the scenario with four nodes in these positions (quantities expressed in meters): $B = (0,0)$, $S = (100, 0)$, $D = (100, 80)$ and $U = (0, 120)$. The $N=4$ positive power levels are $P_i = 0.4\times2^{i-1}\spa mW$, with $i\in\{1,2,3,4\}$, while $P_u=2\spa mW$. In case (a) we have $W = 3$ slots and $\lambda = 0.8212$, whereas in (b) we have $W = 6$ and $\lambda = 1.1918$.}
     \label{fig:opt_multi}
     \vspace{-0.5cm}
\end{figure}

The exact form of the MR policy $\mu_{\lambda}^*(h_d, h_b)$ is obtained by combining the results in Lemmas \ref{lem:binary} and \ref{lem:many_regions}, so for $\lambda\geq1$ we have % (and derived in details in out technical report~\cite{ourtechrep}), is
\begin{equation}
 \mu_{\lambda}^*(h_d,h_b) \!\!=\!\! \left\{
	    \begin{array}{ll}
             \!\!\!0 & \!\!\text{if } h_b < \tilde{g}_0(h_d) \\
             \!\!\!1 & \!\!\text{if } \tilde{g}_0(h_d) < h_b < g_1^+(h_d) \\
             \!\!\!i & \!\!\text{if } g_{i-1}^+(h_d)\! <\! h_b\! <\! g_i^+(h_d)\text{, }\! \text{for } 2\leq i<N \\
             \!\!\!N & \!\!\text{if } h_b > g_{N-1}^+(h_d) \; ,
            \end{array}
            \right.
            \label{optpol_gen_lowk}
\end{equation}
while for $\lambda<1$ we obtain
\begin{equation}
 \mu_{\lambda}^*(h_d,h_b) \!\!=\!\! \left\{
 \begin{array}{ll}
  \!\!\!0 & \!\!\text{if } h_d > \tilde{f}_0(h_b) \\
  \!\!\!N & \!\!\text{if } f_{N-1}^+(h_b) < h_d < \tilde{f}_0(h_b)\\
  \!\!\!i & \!\!\text{if } f_{i-1}^+(h_b)\! <\! h_d\! < \!f_i^+(h_b) \text{, }\! \text{for } 2\leq i<N\\
  \!\!\!1 & \!\!\text{if } h_d < f_1^+(h_b) \; .
  \label{optpol_gen_highk}
 \end{array}
\right.
\end{equation}
A graphical representation of the MR policy in both cases is reported in Figs.~\ref{fig:opt_multi}-(a) and~\ref{fig:opt_multi}-(b), for $\lambda\geq1$ and $\lambda<1$, respectively.

\subsection{Derivation of the optimal $\lambda$}
\label{sec:optlam}
In the previous subsection, we have determined the expression of the MR power allocation strategy $\mu_{\lambda}^*$ for any given value of the blockage weight $\lambda$. The corresponding throughput values attained by the DUE and the CUE when this strategy is adopted are given in the two following propositions.
\begin{prop}
 The expected throughput $\tau(\lambda)$ of a DUE using the MR power allocation strategy $\mu_{\lambda}^*$ is
 \begin{equation}
  \tau(\lambda) = \frac{\pdel}{1+W\pblo}\;,
  \label{prothro}
 \end{equation}
where $W$ is the blockage duration in slots, while $\pdel$ is the expected decoding probability and $\pblo$ is the expected blockage probability, defined as
\begin{eqnarray}
 \pdel & = & \sum_{i=1}^N\int_{\mathcal{Z}_i}p_i(x)e^{-(x+y)}\de x\de y\;,\\
 \pblo & = & \sum_{i=1}^N\int_{\mathcal{Z}_i}q_i(y)e^{-(x+y)}\de x\de y\;.
\end{eqnarray}
\label{proptau}
\end{prop}
\begin{proof}
See Appendix~\ref{app:prooftau}.
\end{proof}

\begin{prop}
 The expected throughput $\sigma(\lambda)$ of the a CUE sharing the channel with a DUE that employs the MR power allocation strategy $\mu_{\lambda}^*$ is
\begin{equation}
 \sigma(\lambda) = \begin{cases} 
                    \displaystyle \frac{e^{\frac{1}{\gamma_{UD}}}\lambda^{\frac{2^{N-1}\gamma_{SD}}{\theta\gamma_{UD}}}\left(e^{-\frac{\theta}{\gamma_{UB}}}-1\right)}{1+\pblo W} + & \\
                    \displaystyle+\frac{1 - \pblo + \pblo We^{-\frac{\theta}{\gamma_{UB}}}}{1+\pblo W}&\!\!\!\!\!\!\!\! \text{if } \lambda<e^{-\frac{\theta}{2^{N-1}\gamma_{SD}}} \\
                    \displaystyle e^{-\frac{\theta}{\gamma_{UB}}}-\frac{\pblo}{1 + \pblo W} &\!\!\!\!\!\!\!\! \text{if } \lambda>e^{-\frac{\theta}{2^{N-1}\gamma_{SD}}}.
                   \end{cases}
\end{equation}
\label{propsigma}
\end{prop}
\begin{proof}
 See Appendix \ref{app:proofsig}.
\end{proof}
Correspondingly, we define the optimal value $\lambda^* = \arg\max_{\lambda}\tau(\lambda)$ as the one which maximizes the DUE expected throughput $\tau(\lambda)$.

\subsection{Single Power Level Scenario}
\label{sec:sinpow}
The model described in the previous sections allows the derivation of the optimal $\lambda$ for any number of possible power levels $N$.
As a case study, we here derive explicitly the computation of the optimal $\lambda$ when $N=1$, which corresponds to an on/off power control where a D2D source either transmit (with fixed power $P_S$) or remains silent.
In this scenario, only two regions ($\mathcal{Z}_0$ and $\mathcal{Z}_1$) exist, and their boundary is the function $\tilde{g}_0(h_d)$ defined in (\ref{defA0}), or, equivalently, its inverse $\tilde{f}_0(h_b)$.
Correspondingly, the throughput is computed as
\begin{equation}
 % \tau(\lambda) & = & \frac{\int_{\mathcal{Z}_1}p_1(x)e^{-(x+y)}\de x\de y}{1+W\int_{\mathcal{Z}_1}q_1(y)e^{-(x+y)}\de x\de y}\nonumber \\
 \tau(\lambda) = \frac{\int_0^{+\infty}\int_{\tilde{g}_0(x)}^{+\infty}p_1(x)e^{-(x+y)}\de y\de x}{1+W\int_0^{+\infty}\int_{\tilde{g}_0(x)}^{+\infty}q_1(y)e^{-(x+y)}\de y\de x}\;.
 \label{optthro_sin}
\end{equation}
The resulting expression is a function of the topology and of the transmit power values $P_S$ and $P_U$ of the DUE and the CUE, respectively. These values can be set in order to attain some predefined SNR values at the intended receiver.
Analogously to the transmissions from $U$ to the BS $B$, which have a target SNR equal to $\rho$, we set a target SNR $\xi$ for the transmission from $S$ to $D$.
This implies that $P_U = \rho d_{UB}^{\alpha}N_0$, while $P_S = \xi d_{SD}^{\alpha}N_0$. By plugging these results into (\ref{mulpi}) and (\ref{mulqi}), we obtain that $\tau(\lambda) = \tau_l(\lambda)$ for $\lambda < e^{-\frac{\theta}{\xi}}$, and $\tau(\lambda) = \tau_k(\lambda)$ for $\lambda \geq e^{-\frac{\theta}{\xi}}$, with
\begin{eqnarray}
 \tau_l(\lambda)& = & \frac{N_1-N_2\lambda^{\frac{1}{z_2}+1}}{N_3-N_4\lambda^{\frac{1}{z_2}}}
 \label{eq:tauhk}
\end{eqnarray}
and
\begin{equation}
 \tau_h(\lambda) = \frac{\displaystyle \lambda}{\displaystyle\left(1+z_2+z_1z_2\right) e^{\frac{\theta}{\rho}}e^{\frac{\theta}{\xi}(z_1+1)}\lambda^{z_1+1} +\frac{z_1W}{1+z_1}}\;,
\label{eq:taulk}
\end{equation}
where $z_1 = \theta\gamma_{SB}/\rho$, $z_2 = \theta\gamma_{UD}/\xi$, while
\begin{eqnarray}
 N_1 &\!\!\!\! = &\!\!\!\! \frac{e^{-\frac{\theta}{\xi}}}{1+z_2}, \quad\quad N_3 = 1 + W\!\left(1-\frac{e^{-\frac{\theta}{\rho}}}{1+z_1}\right), \nonumber
\end{eqnarray}
\begin{eqnarray}
 N_2 &\!\!\!\! = &\!\!\!\! \frac{e^{\frac{1}{\gamma_{UD}}}}{1+z_2}\left(1-\frac{1+z_2}{1+z_2+z_1z_2}e^{-\frac{\theta}{\rho}}\right), \nonumber\\
 N_4 &\!\!\!\! = &\!\!\!\! We^{\frac{1}{\gamma_{UD}}}\left(1-\frac{1+z_2}{1+z_2+z_1z_2}e^{-\frac{\theta}{\rho}}\right).\nonumber
\end{eqnarray}

The function $\tau(\lambda)$ is continuous over all its domain $\mathbb{R}^+$, and its global maximum corresponds to the optimal value $\lambda^*$.
We first analytically compute that $\tau_h(\lambda)$ has its unique maximum at $\lambda_M$, expressed as
\begin{equation}
 \lambda_M = e^{-\frac{\theta}{\xi}}\left(\frac{We^{-\frac{\theta}{\rho}}}{(1+z_1)(1+z_2+z_1z_2)}\right)^{\frac{1}{z_1+1}}\;,
 \label{defKM}
\end{equation}
with $\tau_h(\lambda_M) = \lambda_M/W$.
We then observe that if $W>e^{\frac{\theta}{\rho}}(1+z_1)(1+z_2+z_1z_2)$, then $\lambda_M > e^{-\frac{\theta}{\xi}}$, and the optimal value is $\lambda^*=\lambda_M$. Otherwise, $\tau_h(\lambda)$ is monotonically decreasing for $\lambda\geq e^{-\frac{\theta}{\xi}}$, and $\lambda^*$ is hence equal to the maximum of $\tau_l(\lambda)$ in the interval $(0,e^{-\frac{\theta}{\xi}})$. In this case, we can
numerically find the solution by taking the derivative. The value of $\lambda^*$ is the unique solution of
\begin{equation}
 \lambda^{\frac{1}{z_2}+1} - \frac{N_3}{N_4}\frac{1+z_2}{z_2}\lambda + \frac{N_1}{z_2N_2} = 0
 \label{numK}
\end{equation}
in the interval $(0, e^{-\frac{\theta}{\xi}})$.

\subsection{Discussion and extensions to realistic scenarios}
The mathematical derivations in this section are based on (\ref{mulpi}) and (\ref{mulqi}), which hold for a single cell scenario where only one DUE pair is admitted on each channel. This simple scenario has been chosen in order to better highlight the details and the functioning of our approach, but it is possible to extend it to a more general one by choosing the total perceived power as a feedback from $D$ and $B$, instead of $h_d$ and $h_b$. This feedback would hence include also the interference from the CUEs of surrounding cells, which transmit continuously. The less predictable interference from other DUEs in the same or in an adjacent cell cannot be directly measured, 
but two correcting terms could be added by $S$ to the feedback received from $B$ and $D$, accounting for the extra expected interference from other D2D communications. 
These terms should be tuned in real time, based on the observed throughput. Notice that these modifications would not significantly alter the mathematical derivations presented in this section.

Additionally, the channel condition in realistic scenarios is often characterized by time correlation. It is possible to implement our scheme with time correlated fading coefficients as well. 
A higher performance may be attained, since the CSI collected in a time slot might allow to better predict the required power allocation in the subsequent ones. In this scenario, closed form mathematical expressions should be substituted by numeric quantized solutions, but the rationale of our approach would not be altered. We leave this investigation as a promising research direction for future work.

\section{Mode and Channel Selection}
\label{sec:modchasel}
Differently from the power selection, in our system the mode and channel selection is performed in a centralized manner. The reason, as explained in Sec.~\ref{sec:sysmodel}, lies in the different time scale of these processes and in the amount of information required.
We have proven in Sec.~\ref{sec:optlam} that the expected throughput $\tau(\lambda)$ (of a DUE $S$) and $\sigma(\lambda)$ (of a CUE $U$) sharing the same uplink channel depend only on static information: the fading statistics and the euclidean distances among $S$, $D$, $U$ and the BS $B$. This information is hence also available at the BS.
Assuming that each DUE exploits the optimal power allocation strategy, corresponding to $\lambda=\lambda^*$, the BS can thus also compute the expected throughput for $S$ and $U$ if the D2D mode is chosen. The throughput achievable via D2B mode is instead equal to $\bar{\tau} = 0.5e^{-\frac{\theta}{\rho}}$ for both\footnote{For DUE $S$, we assume that the BS relays its data to $D$ in a full-duplex manner, and that the bottleneck lies in the uplink.} $U$ and $S$. This comes from the fact that, when D2B mode is used, both terminals set their power to achieve a target SNR equal to $\rho$, no cross-tier interference is experienced, and a TDMA scheme is adopted, as stated in Sec.~\ref{sec:sysmodel}.

The channel and mode selection is hence performed in a joint manner. Assuming $K$ CUEs and $N$ DUE pairs in the cell,
the BS first fills a $K\times N$ matrix $\mathbf{T}$ with the throughput achievable by any possible CUE-DUE couple sharing the same channel. For each couple $(U_i, S_j)$, with $i\in\{1,2,\ldots K\}$ and $j\in\{1,2,\ldots N\}$, the BS first computes the optimal $\lambda^*$ to be used in D2D mode and the corresponding expected throughput values $\tau(\lambda^*)$ and $\sigma(\lambda^*)$ for DUE $S_j$ and CUE $U_i$, respectively. Subsequently, it sets
\begin{equation}
 \mathbf{T}(i, j) = \begin{cases}
                     \tau(\lambda^*) + \sigma(\lambda^*) & \text{if } \tau(\lambda^*) > \bar{\tau} \\
                     2\bar{\tau} = e^{-\frac{\theta}{\rho}} & \text{otherwise}\;.
                    \end{cases}
 \label{match}
\end{equation}
In the former case, D2D mode will be chosen if a channel is assigned to (and hence shared by) $U_i$ and $S_j$; in the latter, instead, D2B will be chosen. Notice that this mode selection ensures that D2D communications are leveraged only when they are advantageous.

Once the entire matrix $\mathbf{T}$ is filled, the channel selection is to be performed by choosing how to pair the CUEs and DUEs in order to attain the highest overall throughput. This can be done by considering a bipartite weighted graph where the two sets of vertices are the CUEs and the DUEs. The weight of the edge between CUE $U_i$ and DUE $S_j$ is given by $\mathbf{T}(i,j)$.
The optimal channel allocation is found as the maximum weighted matching on the graph, and in our case it is obtained using the Hungarian method~\cite{hungarian}, which solves the problem in polynomial time~\cite{N19}.
Referring to the pairing vector $\mathbf{\Lambda}$ and the mode vector $\mathbf{x}$ in (\ref{probform2}), for any edge $(i,j)$ included in the obtained matching, we set $\mathbf{\Lambda}(i) = D_j$, while $\mathbf{x}(i)$ is set to 1 or 0 according to the preferred mode for the spectrum sharing between $U_i$ and $D_j$, as indicated by (\ref{match}).

\section{Performance Evaluation}
\label{sec:results}
In this section, we measure the performance of our channel/mode and power selection scheme. We focus on a single-cell scenario, with $K=5$ CUEs (equal to the number of considered uplink channels) and $M=5$ DUE pairs.
We set the cellular radius to $d_R=200\spa\rm{m}$, the maximum length of a D2D link to $d_L=100\spa\rm{m}$, the path loss exponent to $\alpha = 4$, the CUE target SNR at $B$ to $\rho=0\spa\rm{dB}$, with $N_0=-90\spa\rm{dBm}$, and the decoding threshold to $\theta = 0\spa\rm{dB}$.

We first analyze the scenario where a single power level is available at the DUE source, meaning that an on/off power allocation is performed. The power level is set at the beginning of each epoch in order to have a target SNR equal to $\xi$, according to Sec.~\ref{sec:sinpow}. In general, $\xi$ could be set autonomously by each DUE source, but in this work, we consider it to be equal for all the DUE pairs.

Our investigation is focused on the tradeoff between the two tunable parameters $W$ and $\xi$, since they have the highest impact on the overall cell throughput. The considered metrics are the average CUE throughput $\Omega_C$ and the average DUE throughput $\Omega_D$.
We named CMP our channel/mode/power selection approach, and we compare it with an adapted version of the scheme proposed in~\cite{modpowcon}, which we rename here as GEO. In this scheme, the mode selection is based only on a geographic basis: a DUE $S$ selects the D2D mode if $\kappa d_{SD}^{-\alpha} \leq d_{SB}^{-\alpha}$, where $d_{SD}$ is the D2D link length, $d_{SB}$ is the distance of $S$ from the BS, and $\kappa\geq0$ is a tunable parameter which regulates the tradeoff between D2D and D2B mode.
The same target SNR $\rho$ is used in both modes (thus setting the transmit power accordingly), but no blockage mechanism is employed. As to the channel selection, we adopt the same mechanism used for CMP, thus letting the BS computing the expected throughput of any possible CUE-DUE pair and selecting the best allocation through the Hungarian algorithm. In this paper, we set $\kappa = 0.8$.

The simulations are performed with a MATLAB simulator. All the results are obtained by averaging over 1000 random topologies and considering a single epoch of $T_e=100$ time slots. Each topology is randomly generated according to the cellular radius and the maximum D2D link length.
In each time slot, the channel fading coefficients are generated from an exponential random variable of unitary mean. The transmission of feedback and control packets
is assumed to be error-free. The values of $h_d$ and $h_b$ computed at each time slot have been quantized with step 0.01 and restricted to the interval $[0,5]$.

\subsection{Average CUE and DUE Throughput}
\begin{figure}
    \centering
    \includegraphics[width=\figL]{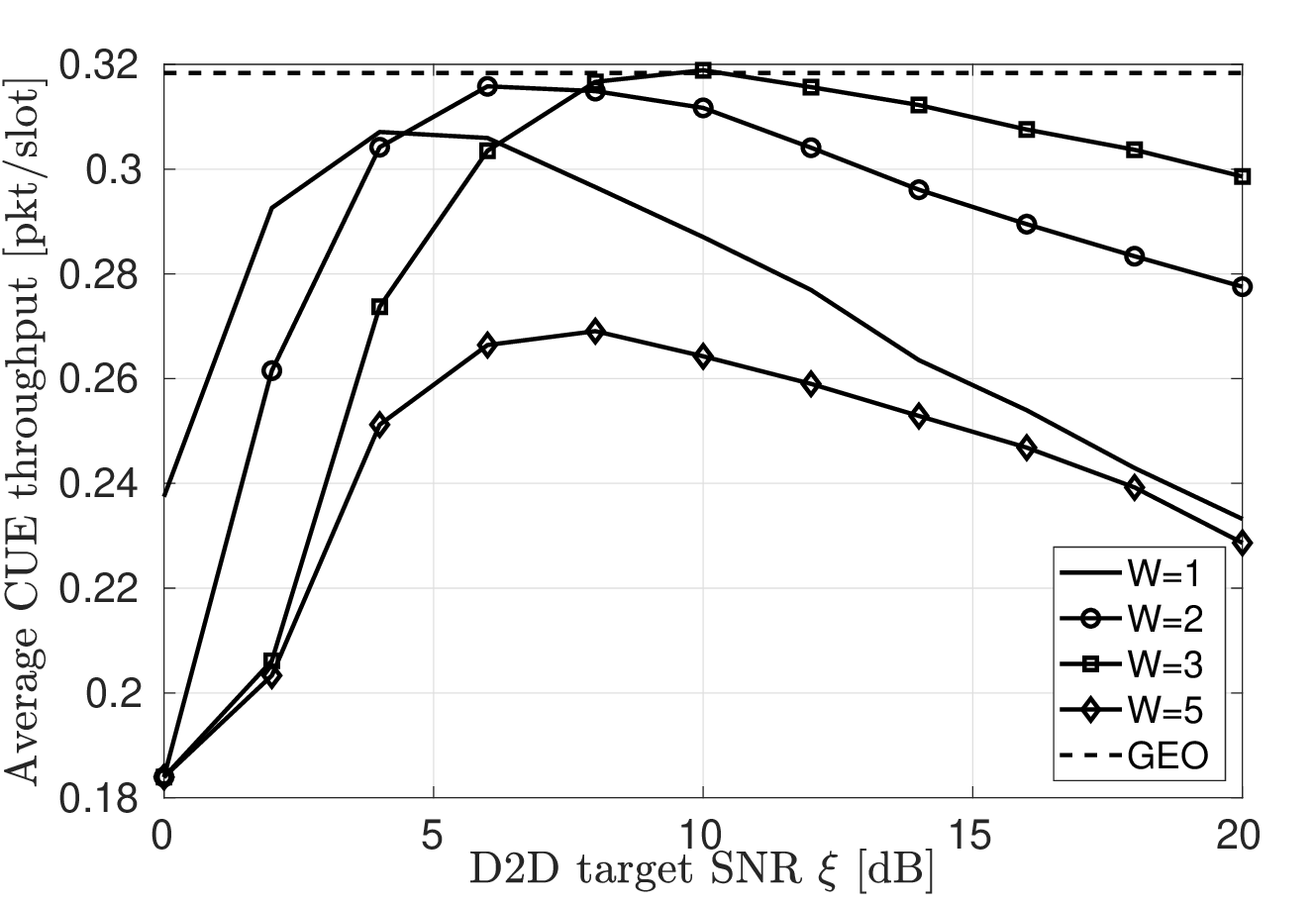}
     \caption{\small The average CUE throughput $\Omega_C$, as a function of the D2D target SNR $\xi$.}
     \label{fig:throU_vs_mu}
     \vspace{-0.5cm}
\end{figure}

In Fig.~\ref{fig:throU_vs_mu}, we depict $\Omega_C$, the average CUE throughput, as a function of the D2D target SNR $\xi$ and for different values of $W$. For CMP, we observe that for each value of $W$ there is a value of $\xi$ for which $\Omega_C$ is maximized. For higher values of $\xi$, the interference from the DUE becomes significant. For lower values of $\xi$ instead, most DUEs are likely to opt for the D2B mode, thus halving the spectrum resources of the DUEs. 

On the other side, if the value of $\xi$ is known, there exists an optimal value of $W$ to maximize $\Omega_C$.
A too low value of $W$ makes it convenient for DUEs to transmit in D2D mode even if this impairs the CUE communications, while a too high value of $W$ will make the D2B mode more effective, leading to DUEs exclusively using half of the resources. We observe also that the optimal value for $W$ increases as $\xi$ increases, since a higher $\xi$ means a higher power for the DUEs, which also means a higher disturbance for the CUEs.
Finally, we observe that $\Omega_C$ is always higher for the GEO strategy. This is not surprising, since the GEO is designed to protect the CUE transmissions from any cross-tier interference.
Notice, however, that for $W=3$ and $\xi=10\spa\rm dB$, CMP is still able to grant the same average CUE throughput as GEO.
\begin{figure}
    \centering
    {\includegraphics[width=\figL]{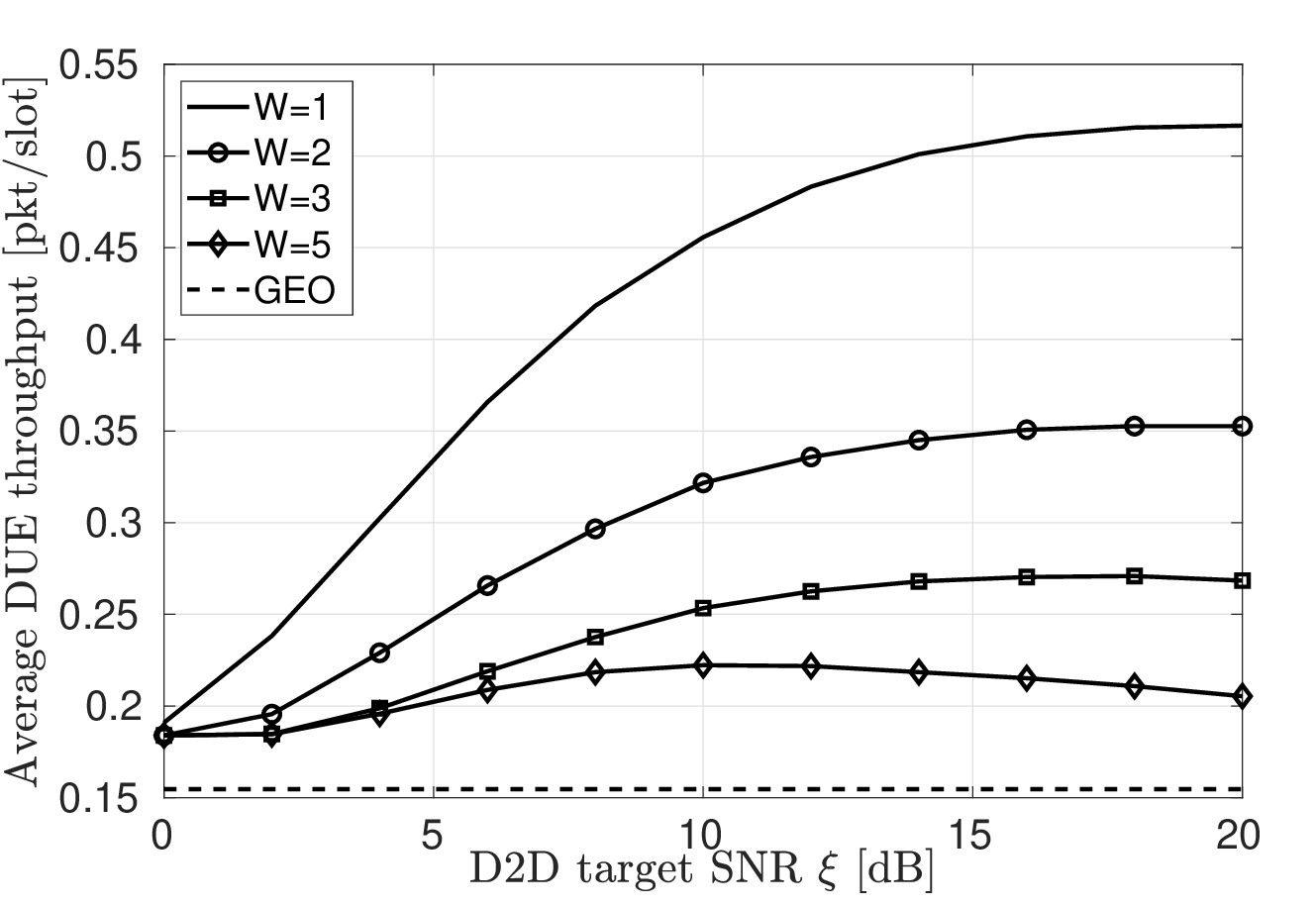}
     \caption{\small The average DUE throughput $\Omega_D$, as a function of the D2D target SNR $\xi$.}
     \label{fig:throS_vs_mu}
     }
     \vspace{-0.4cm}
\end{figure}

The value of $\Omega_D$, the average DUE throughput, is shown in Fig.~\ref{fig:throS_vs_mu}.
We observe that, in the CMP case, it is convenient for DUEs to increase the value of $\xi$, at least for low values of the blockage duration, $W \leq 4$. In any case, the value of $\xi$ can not be arbitrarily increased, otherwise a blockage is triggered after every D2D transmission.
We also notice that $\Omega_D$ is much higher in the case of CMP than GEO, for any of the considered values of $W$ and $\xi$. This comes at the cost of a reduced CUE throughput, as observed in Fig.~\ref{fig:throU_vs_mu}.
\begin{figure}
    \centering
     {\includegraphics[width=\figL]{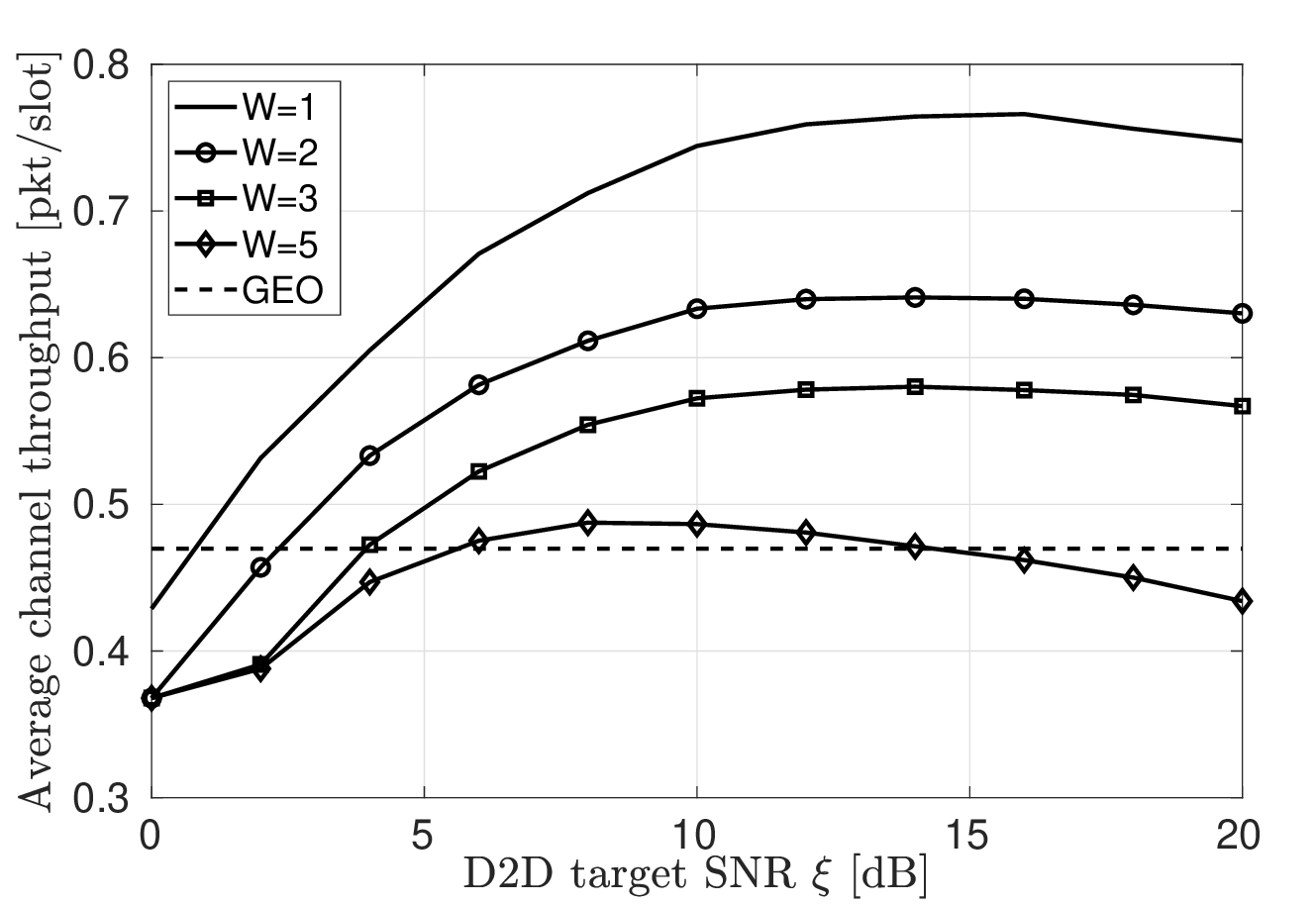}
     }
     \caption{\small The channel throughput $\Omega_{C+D}$, as a function of the D2D target SNR $\xi$.}
     \label{fig:throAll_vs_mu}
     \vspace{-0.4cm}
\end{figure}

In order to analyze the cost-benefit balance for CMP, we depict in Fig.~\ref{fig:throAll_vs_mu} the average channel throughput, $\Omega_{C+D} = \Omega_C+\Omega_D$. In general, the maximum for the system throughput $\Omega_{C+D}$ is obtained for $W=1$, and for each value of $W$ the performance are maximized for one finite value of $\xi$. 
The total throughput increase for CMP with respect to GEO is particularly significant for low values of $W$ and $\xi >4\spa\rm{dB}$.
In particular, for $W=1$ and $\xi =16\spa\rm{dB}$, CMP outperforms GEO by about $63\%$ in terms of total throughput.

This significant increase in $\Omega_{C+D}$ comes at the cost of a consistent decrease in $\Omega_{U}$, the CUE throughput.
Fairness also plays an important role, and we can find the optimal parameter setup by comparing $\Omega_C$ and $\Omega_D$ in the previous pictures. For values $W\geq3$, we have $\Omega_C>\Omega_D$ for every choice of $\xi$. Conversely, with $W=1$ and $W=2$ we can better balance the available resources, and even get the same throughput for CUEs and DUEs.
This happens for $\xi=4\spa\rm dB$ when $W=1$, and for $\xi=9\spa\rm dB$ when $W=2$. In both cases, we have approximately $\Omega_C=\Omega_D=0.31\rm pkt/slot$, and hence $\Omega_{C+D} = 0.62\rm pkt/slot$. By using GEO, we would still get the same value for the CUE throughput, but only a halved value for $\Omega_D$, resulting in $\Omega_{C+D}=0.46\rm pkt/slot$.
In case of no spectrum sharing at all, the CUEs and DUEs should always orthogonally share the channels, with a total channel throughput of $0.37\spa\rm{pkt/slot}$.

In other words, it is possible to tune the CMP parameters to achieve the maximum fairness. Even in this case, the relative gain in terms of system throughput over GEO is of about $35\%$, while over the case of no D2D transmission the relative gain is about $68\%$.

\subsection{Multiple power levels}
We now extend the analysis to the case where multiple power levels $N>1$ can be dynamically selected.
\begin{figure}
    \centering
    {\includegraphics[width=\figL]{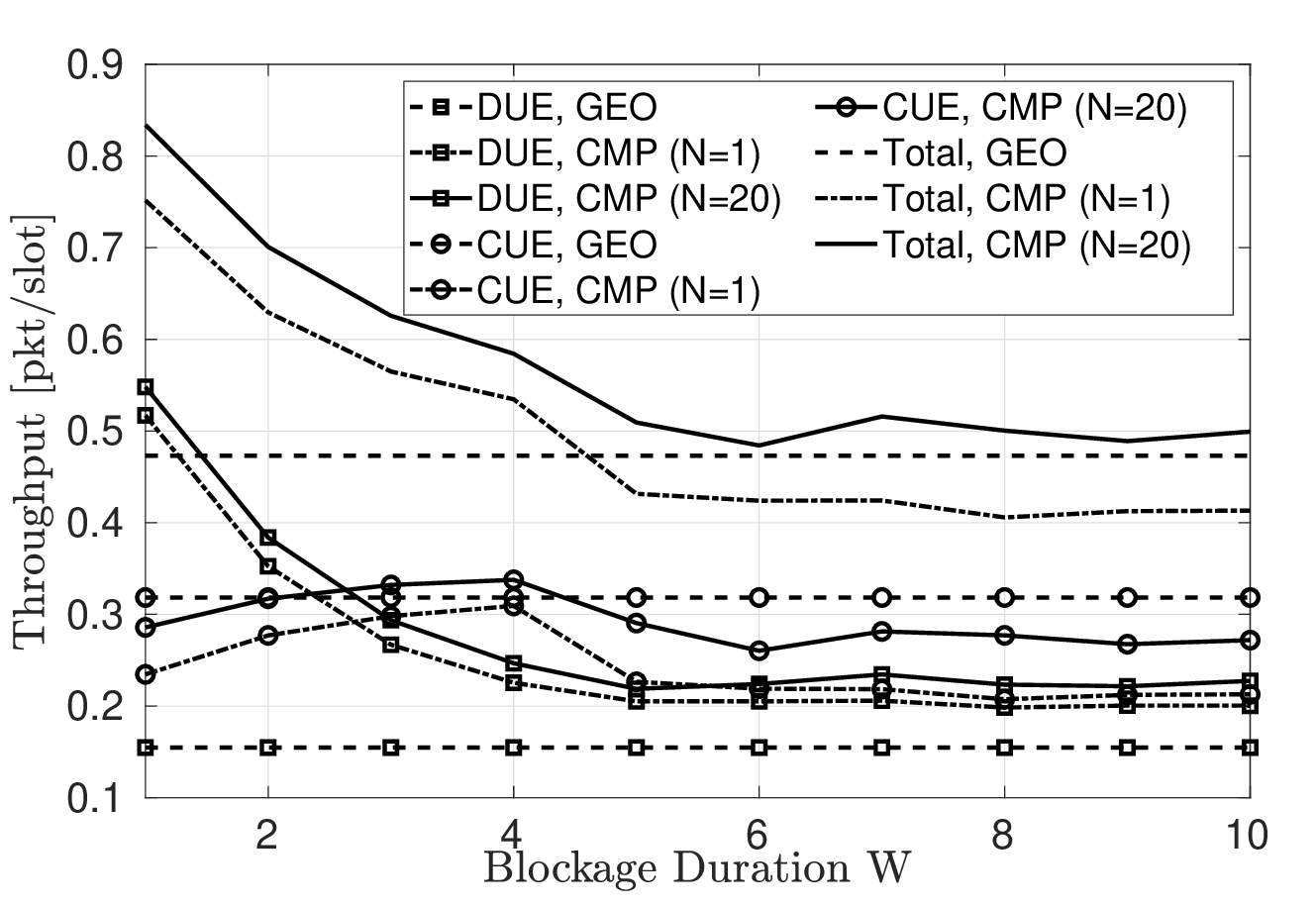}
     \caption{\small The CUE, DUE and channel expected throughput, as a function of the blockage duration $W$, for GEO, CMP with $N=1$ ($\xi=20\spa\rm{dB}$) and CMP with $N=20$.}
     \label{fig:Multi_ThroSeU_vs_W}
     }
     \vspace{-0.5cm}
\end{figure}
In this case, we assume that $N=20$ logarithmically spaced power levels are available, with the highest one being equal to the maximum device transmit power $P_m=200\spa\rm mW$: $\mathcal{P} = P_m2^{-i}$, for $i\in\{0,1,2,\ldots, N-1\}$. Differently from the case with $N=1$, where the single available power level was set depending on the target SNR $\xi$ and on the topology, here we move to a more realistic setup, where the $N$ transmit power levels are fixed, and equal for all the DUEs, irrespective of their location.

In Fig.~\ref{fig:Multi_ThroSeU_vs_W}, we plot $\Omega_C$ and $\Omega_D$ for GEO, CMP with $N=1$ (and $\xi = 20\spa\rm{dB}$), and CMP with $N=20$, as a function of the blockage duration $W$.
Even if the multiple power levels are fixed, and not depending on the topology, when $N=20$ CMP is able to increase its CUE throughput by up to 10\% with respect to the case $N=1$. This is due to the fact that when channel conditions are good, a lower transmission power can be used, hence reducing the cross-tier interference while still allowing to deliver data packets.
Correspondingly, also $\Omega_D$ is increased, resulting in a total channel throughput increase of approximately 0.06 pkt/slot. Notice that, for $W=3$ and $W=4$, both the CUE and the DUE throughput attained by CMP (with $N=20$) are higher than those granted by GEO.
If fairness is to be pursued by selecting the value of $W$ which makes $\Omega_C=\Omega_D$, the optimal value\footnote{Fractional values for $W$ can be obtained by considering it as a random variable with a properly set mean value.} is around 2.5. Under these conditions, CMP with $N=20$ can achieve $\Omega_{C+D} = 0.65\spa\rm pkt/slot$, which is about 10\% higher than in the single power level case, and four times the one offered by GEO.

\section{Conclusions}
\label{sec:conclu}
In this paper, we presented a novel channel/mode and power selection scheme for D2D communications over uplink channels. We proposed to split the resource allocation, making centralized decisions based only on static or stochastic information, and locally exploiting fast changing information by means of a proper power allocation strategy, implemented in a distributed fashion.
We theoretically derived the optimal power allocation strategy to be implemented in order to maximize the D2D communication throughput. For the case of a single available power level, we also obtained the explicit expressions for both the DUE and the CUE throughput.
The obtained results confirmed the effectiveness of our approach: without the need for gathering all the information at a centralized entity, which would imply a huge amount of overhead, we showed the benefits attained by distributedly exploiting local information, in terms of throughput and fairness.
In a future work, we plan to extend the analysis to more complex scenarios, where inter-cell interference is also modeled, and multi-hop communications can be established as well.

\appendices

\section{Mathematical proofs}
\subsection{Proof of Lemma \ref{lem:many_regions}}
\label{app:prooflem}
Let us call $G_i(h_d,h_b) = p_i(h_d)-\lambda q_i(h_b)$ the expected gain for state $(h_d,h_b)$ when power level $i$ is selected.
The state $(h_d,h_b)$ belongs to $\mathcal{Z}_i$ if $G_i(h_d,h_b)>G_j(h_d,h_b)$, $\forall j\neq i$. Since $G_i(h_d,h_b)$ is continuous $\forall i$, if there exists a boundary $\mathcal{Q}\subset\mathcal{Z}$ between $\mathcal{Z}_i$ and $\mathcal{Z}_{i+s}$, with $s\in\mathbb{N}$ and $s>1$, it would follow that, for any $(h_d,h_b)\in\mathcal{Q}$, $G_i(h_d,h_b) = G_{i+s}(h_d,h_b) > G_j(h_d,h_b)$, $\forall j\neq i,i+s$. In particular, if we chose $j = i+1$, we would have $G_i(h_d,h_b) = G_{i+s}(h_d,h_b) > G_{i+1}(h_d,h_b)$.

However, this is not possible. In fact, consider the function $v_{h_d,h_b}(x)$, defined as $p_x(h_d)-\lambda q_x(h_b)$, where now $x\in\mathbb{R}^+$, while $h_d$ and $h_b$ are fixed. Clearly, for $x\in\{0,1,2,\ldots,N\}$, we find $v_{h_d,h_b}(x)=G_x(h_d,h_b)$. This function has no minimum points in $\mathbb{R}^+$. Depending on the given values of $h_d$ and $h_b$, it either has a single maximum in
\begin{equation}
 x_M = 1 + \log_2\left(\frac{\ln(p_1(h_d)) - \ln(q_1(h_b))}{\ln\ln(p_1(h_d)) - \ln\ln(q_1(h_b))+\ln(\lambda)}\right)
\end{equation}
or it is strictly monotonic over all its domain.

It follows that if $G_i(h_d,h_b) = G_{i+s}(h_d,h_b)$, then $x_M$ exists in the interval $(i, i+s)$; moreover, $v_{h_d,h_b}(x)$ is greater than both $G_i(h_d,h_b)$ and $G_{i+s}(h_d,h_b)$ for any $x\in(i, i+s)$, and therefore also for $x = i+1$, which contradicts the hypothesis. This proves the lemma.

\subsection{Proof of Theorem \ref{teo:optsol_multiple}}
\label{app:proofteo}
The boundary between $\mathcal{Z}_i$ and $\mathcal{Z}_{i+1}$ is by definition $g_i^+(h_d)$ (if $\lambda\geq1$), or $f_i^+(h_b)$ (if $\lambda<1$). Henceforth, these curves, together with $\tilde{g}_0(h_d)$, are the only admissible boundaries between the regions $\mathcal{Z}_i$ for $i=0,\dots,N$.
According to Remark \ref{rem:noint}, these $N$ curves never intersect each other. Since they are continuous functions of either $h_d$ (if $\lambda\geq1$) or $h_b$ (if $\lambda<1$), it follows that they divide $\mathcal{Z}$ into at most $N+1$ regions. The number of regions can however be lower. In fact, while it is always $g_i^+(h_d)\cap\mathcal{Z}\neq\emptyset$, $\forall i\in\{1,2,\ldots,N\}$, as per Remark \ref{rem:exig}, the same does not hold for the curves $f_i^+(h_b)$.
It can be shown that if $\gamma_{SD}2^i<\theta/\ln2$, then $f_i^+(h_b)<0$, $\forall h_b\in\mathbb{R}^+$. This means that
the entire curve lies outside the region $\mathcal{Z}$. In this case, for every $j$ such that $0<j\leq i$ no region $\mathcal{Z}_j$ exists,  due to Remark \ref{rem:order}.
In the extreme case, if it is $f_{N-1}^+(h_b)<0$, $\forall h_b$, then $\mathcal{Z}$ is divided into 2 regions, namely $\mathcal{Z}_0$ and $\mathcal{Z}_N$, by the curve $\tilde{f}_0(h_b)$.

\subsection{Proof of Proposition \ref{proptau}}
\label{app:prooftau}
Due to the time uncorrelation of the fading coefficients, we can use Renewal Theory to compute, for a given value of $\lambda$, the expected throughput $\tau(\lambda)$ of the corresponding MR strategy $\mu_{\lambda}^*$. The throughput is equal to the ratio between the expected packet decoding probability and the expected time between two subsequent transmissions.
Let us call $\chitx$ the event of having a favourable condition to transmit according to strategy $\mu_{\lambda}^*$, that is, the event of being in a state $(h_d, h_b)\notin\mathcal{Z}_0$.
The probability of this event is $\ptx = \mathbb{P}[\chitx] = 1 - \int_{\mathcal{Z}_0}e^{-(x+y)}\de x\de y$.
The throughput is hence given by
\begin{equation}
 \tau(\lambda) = \frac{\mathbb{E}[p_{\mu_{\lambda}^*(h_d,h_b)}(h_d)|\chitx]}{W\mathbb{E}[q_{\mu_{\lambda}^*(h_d,h_b)}(h_b)|\chitx] + 1/\ptx}\;,
 \label{thro}
\end{equation}
where $\mathbb{E}[p_{\mu_{\lambda}^*(h_d,h_b)}(h_d)|\chitx]$ is the expected decoding probability of strategy $\mu_{\lambda}^*$ given that a transmission is performed. Similarly, $\mathbb{E}[q_{\mu_{\lambda}^*(h_d,h_b)}(h_b)|\chitx]$ is the expected blockage probability of strategy $\mu_{\lambda}^*$ given that a transmission is performed. At the denominator, the average time between two subsequent transmissions is hence the sum of the expected blockage duration and the expected time $1/\ptx$ to be waited before the conditions for a transmission are met.

From the expression of $\mu_{\lambda}^*$ in (\ref{optpol_gen_lowk}) and (\ref{optpol_gen_highk}), we have that $\mathbb{E}[p_{\mu_{\lambda}^*(h_d,h_b)}(h_d)|\chitx] = \pdel/\ptx$, while $\mathbb{E}[q_{\mu_{\lambda}^*(h_d,h_b)}(h_b)|\chitx] = \pblo/\ptx$,
which can be plugged into (\ref{thro}) to obtain (\ref{prothro}).

\subsection{Proof of Proposition \ref{propsigma}}
\label{app:proofsig}
The considered system evolves in time through \emph{blockage phases} (B phases), when the DUE $S$ is forced to be silent, alternated with \emph{transmission phases} (T phases), when $S$ is allowed to tramsmit. We call $\delta(\rm B)$ and $\delta(\rm T)$ the expected durations of a B and a T phase, respectively. According to Renewal Theory, the expected throughput $\sigma(\lambda)$ is obtained as
\begin{equation}
 \sigma(\lambda) = \frac{\sigma(\lambda|\rm T)\delta(T) + \sigma(\lambda|B)\delta(\rm B)}{\delta(\rm T)+\delta(\rm B)}\;,
 \label{rentheo}
\end{equation}
where $\sigma(\lambda|\rm{X})$ is the expected throughput in phase $\rm X\in\{T,B\}$.

The duration of a B phase is fixed, $\delta({\rm B})=W$. The duration of a T phase is instead stochastic. Since the fading coefficients are time uncorrelated, the probability of triggering a blockage is $\pblo$ at every time slot, and $\delta({\rm T})=1/\pblo$. 
The throughput $\sigma(\lambda|\rm{B})$ is simply equal to $e^{-\theta/\rho}$, since no intra-cell interference is present in a B phase.
During a phase T, the last slot has zero throughput, since in this slot a blockage is triggered, meaning that the SINR falled below the decoding threshold $\theta$. Therefore, $\sigma(\lambda|\rm{T}) = (\delta(\rm{T})-1)/\delta(\rm{T})\tilde{\sigma}(\delta|\rm{T})$, where $\tilde{\sigma}(\lambda|\rm{T})$ is the expected throughput during a T phase excluding the last slot.
Equation (\ref{rentheo}) can be rewritten as
\begin{equation}
 \sigma(\lambda) = \frac{\left(\frac{1}{\pblo}-1\right)\tilde{\sigma}(\lambda|\rm T) + We^{-\frac{\theta}{\rho}}}{\frac{1}{\pblo}+W}\;.
 \label{rentheo2}
\end{equation}

In order to find $\tilde{\sigma}(\lambda|\rm{T})$, we condition on the probability that $S$ transmits. We call this event $\chitxT$, and its probability is $\ptxT$, then
\begin{eqnarray}
 \tilde{\sigma}(\lambda|\rm{T}) & = & \tilde{\sigma}(\lambda|\rm{T},\chitxT=1)\ptxT + \nonumber\\
 & & + \tilde{\sigma}(\lambda|\rm{T},\chitxT=0)(1-\ptxT)\;.
\end{eqnarray}
Within a T phase, $S$ transmits when the system state $(h_d,h_b)$ does not belong to $\mathcal{Z}_0$. The probability that the system enters a state $(x,y)\in\mathcal{Z}_i$, with $i>0$, conditioned on the fact that a blockage is not triggered (since we are within a T phase) is equal to $e^{-x}e^{-y}(1-q_i(y))/(1-\pblo)$. Hence
\begin{equation}
 \ptxT = \sum_{i=1}^N\int_{\mathcal{Z}_i}\frac{1-q_i(y)}{1-\pblo}e^{-x}e^{-y}\de x\de y = \frac{\ptx-\pblo}{1-\pblo}\;,
\end{equation}
where $\ptx=\mathbb{P}[\chitx]=\sum_{i=1}^N\int_{\mathcal{Z}_i}e^{-x}e^{-y}\de x\de y$ is the unconditioned transmission probability.
The throughput $\tilde{\sigma}(\lambda|\rm{T},\chitxT=1)$ in a T phase when $S$ transmits is always 1, otherwise a blockage would be triggered, and the system would move into a B phase. If $S$ does not transmit, there is no intra-cell interference, and the throughput is equal to $\mathbb{P}[\rho h_b>\theta|\chitx=0]$. Therefore, $\tilde{\sigma}(\lambda|\rm{T},\chitxT=0)$ corresponds to the probability $\pok$ that $h_b>\theta/\rho$, given that the system status $(h_d,h_b)$ belongs to $\mathcal{Z}_0$. We have
\begin{equation}
 \tilde{\sigma}(\lambda|\rm{T}) = \ptxT + \pok(1-\ptxT)\;.
 \label{inter}
\end{equation}

To compute $\pok$, we distinguish two cases. If $\lambda > e^{-\frac{\theta}{2^{N-1}\gamma_{SD}}}$, $\mathcal{Z}_0$ is delimited by a straight line, since $h^*<0$, as per Lemma \ref{lem:binary}. Correspondingly,
\begin{eqnarray}
 \pok & = & \frac{\int_{\mathcal{Z}_0}e^{-x}e^{-y}\de x\de y - \int_0^{+\infty}\int_0^{\frac{\theta}{\rho}}e^{-x}e^{-y}\de x\de y}{\int_{\mathcal{Z}_0}e^{-x}e^{-y}\de x\de y} = \nonumber \\
 & = & \frac{e^{-\frac{\theta}{\rho}}-\ptx}{1-\ptx}.
\end{eqnarray}

If instead $\lambda < e^{-\frac{\theta}{2^{N-1}\gamma_{SD}}}$, then $\mathcal{Z}_0$ is limited by $\tilde{g}_0(h_d)$, as defined in (\ref{defA0}), and the integral over $h_d$ is computed only for $h_d>h^*$:
\begin{eqnarray}
 \pok & = & \frac{1-\ptx - \int_{h^*}^{+\infty}\int_0^{\frac{\theta}{\rho}}e^{-x}e^{-y}\de x\de y}{1-\ptx}\nonumber \\
 & = & \frac{1-\ptx-e^{\frac{1}{\gamma_{UD}}}\lambda^{\frac{2^{N-1}\gamma_{SD}}{\theta\gamma_{UD}}}(1-e^{-\frac{\theta}{\rho}})}{1-\ptx} \; .
\end{eqnarray}
By plugging the expressions of $\pok$ and $\ptxT$ into (\ref{inter}), and then into (\ref{rentheo2}), the Proposition is proved.

\bibliographystyle{IEEEtran}
\bibliography{IEEEabrv,biblio}

\end{document}